\documentclass[lettersize,journal]{IEEEtran}
\usepackage{algorithm}
\usepackage{algorithmic}
\usepackage{amsfonts}
\usepackage{amsmath}
\usepackage{textcomp}
\usepackage{array}
\usepackage{xcolor}
\usepackage{epsfig}
\PassOptionsToPackage{hyphens}{url}
\usepackage{svg}
\usepackage[font=small, labelfont=bf]{caption}
\usepackage{color}

\usepackage{dingbat}
\usepackage{stfloats}
\usepackage{tabularx}
\usepackage{mdwmath}
\usepackage{multirow}
\usepackage{enumerate}

\usepackage{amssymb}
\usepackage{graphicx}
\usepackage{wasysym}
\usepackage{epstopdf}
\usepackage{subfigure}
\usepackage{mathrsfs}

\usepackage{multirow}
\usepackage{CJKutf8}
\usepackage[utf8]{inputenc}

\usepackage{url}

\usepackage{threeparttable}
\def\BibTeX{{\rm B\kern-.05em{\sc i\kern-.025em b}\kern-.08em
    T\kern-.1667em\lower.7ex\hbox{E}\kern-.125emX}}
\usepackage{balance}
\begin{document}

\title{CRUcialG: Reconstruct Integrated Attack Scenario Graphs by Cyber Threat Intelligence Reports}

\author{Wenrui Cheng,
        Tiantian Zhu*,
        Tieming Chen,
        Qixuan Yuan,
        Jie Ying,
        Hongmei Li,
        Chunlin Xiong,
        Mingda Li,
        Mingqi Lv,
        and Yan Chen, IEEE Fellow
		
}


\maketitle

\def \sn {CRUcialG}

\begin{abstract}
Cyber Threat Intelligence (CTI) reports are factual records compiled by security analysts through their observations of threat events or their own practical experience with attacks. In order to utilize CTI reports for attack detection, existing methods have attempted to map the content of reports onto system-level attack provenance graphs to clearly depict attack procedures. However, existing studies on constructing graphs from CTI reports suffer from problems such as weak natural language processing (NLP) capabilities, discrete and fragmented graphs, and insufficient attack semantic representation. Therefore, we propose a system called \sn{} for the automated reconstruction of attack scenario graphs (ASGs) by CTI reports. First, we use NLP models to extract systematic attack knowledge from CTI reports to form preliminary ASGs. Then, we propose a four-phase attack rationality verification framework from the tactical phase with attack procedure to evaluate the reasonability of ASGs. Finally, we implement the relation repair and phase supplement of ASGs by adopting a serialized graph generation model. We collect a total of 10,607 CTI reports and generate 5,761 complete ASGs. Experimental results on CTI reports from 30 security vendors and DARPA show that the similarity of ASG reconstruction by \sn{} can reach 84.54\%. Compared with SOTA (EXTRACTOR and AttackG), the recall of \sn{} (extraction of real attack events) can reach 88.13\% and 94.46\% respectively, which is 40\% higher than SOTA on average. The F1-score of attack phase verification is able to reach 90.04\%. 
\end{abstract}


\section{Introduction}\label{sec:intro}
\IEEEPARstart{W}{ith} the continuous advancement of information technology, cyber attacks have evolved from traditional single means to complex and changeable advanced persistent threats (APTs). Attackers usually adopt phased and diversified tactical strategies to achieve their attack goals. Fully revealing the attacker’s tactical intentions and combat methods is crucial to attack detection and response. To characterize attack behaviors and acquire attack representations, MITRE has introduced the ATT\&CK \cite{attck} framework, which aggregates known attacks into a structured matrix comprising tactics and techniques. By providing textual descriptions, the framework outlines the potential means by which attackers may carry out specific objectives. However, ATT\&CK presents the tactic, technique, and procedure (TTP) matrix in the form of a knowledge graph while downplaying the procedure information in real attacks (i.e., the “P” information in TTP is just a ranking of tactics), and cannot effectively correlate the contextual content of an attack (i.e., the attack timeline, the origin and target of the attack, and changes in attack phases, etc.). For example, analysts try to improve the attack detection rate by covering as many different techniques in ATT\&CK as possible \cite{milajerdi2019holmes,zhu2023aptshield,hassan2020tactical}, but the single-point technology will bring catastrophic false positives (as mentioned by Rapsheet \cite{hassan2020tactical}, six out of the top ten most commonly used attacker techniques frequently occur in normal user activity), making it difficult to apply in real scenarios. 
\par
In addition, textual data (such as the description of tactics and techniques in ATT\&CK) have limitations. Compared with textual data, attack graphs (e.g., attack scenario graphs or provenance graphs) can not only express attack-related information but also be better processed by machines (a key technique used by reserchers is data provenance \cite{bates2015trustworthy, hassan2020omegalog}, which iteratively parses low-level events into a causal dependency graph that maps out the entire history of system execution). For example, an attack scenario graph (ASG), which consists of nodes (e.g., processes, files, sockets) and edges (e.g., reads, writes, executes), is a machine-processable data structure that enables the reconstruction of an attack scenario from multiple dimensions. It effectively depicts the causality, attack paths, and impact of threat events, providing contextual information about the attack. However, on the one hand, there are relatively few existing data sets containing complete ASGs (e.g., only 18 and 17 ASGs can be extracted from the data set of Darpa Engagement 3 and 4, respectively), and on the other hand, it is difficult to obtain ASG through recurrence attacks (e.g., the ASGs acquisition methods based on sandbox \cite{song2021towards,pan2018flowcog} or terminal probe \cite{shi2020vahunt,qu2017appshield} have the problems of sandbox confrontation and insufficient data granularity).
\par
Fortunately, security analysts write CTI reports through attack events or attack practices. These reports provide the attacker's tactics and techniques, and narrate the attack procedure in chronological order. At present, some research studies have attempted to map the content of CTI reports onto system-level attack provenance graphs (it is at the same level as the ASG mentioned in this article) to clearly depict attack procedures. However, due to the complexity and particularity of CTI reports, existing ASG extraction methods have problems such as weak processing ability of multi-source heterogeneous CTI reports, discrete and fragmented graphs, and insufficient attack semantics, which are difficult to jointly satisfy the requirements of completeness and availability. 

\begin{table*}[h!t]
\caption{Comparison of CTI-based related work on various attack representations. The three rightmost columns assess the universality, connectivity, and rationality of the graph construction system. The solidness of the marked circle reflects the degree: High (\CIRCLE), Medium (\RIGHTcircle), Low (\Circle).}
\label{table:related-work}
\resizebox{\textwidth}{!}{
\centering
\begin{tabular}{|>{\centering\arraybackslash}m{3.5cm}|c|c|>{\centering\arraybackslash}m{3cm}|c|c|c|}
\hline
\textbf{Attack Representation}   & \textbf{System}                 & \textbf{Machine-processable} & \textbf{Attack Procedure Information} & \textbf{Method Universality} & \multicolumn{1}{l|}{\textbf{Graph Connectivity}} & \multicolumn{1}{l|}{\textbf{Graph Rationality}} \\ \hline
Attack Entity        & iACE \cite{liao2016acing}                            & \CIRCLE  & \Circle                & \textbackslash  & \textbackslash                                 &  \textbackslash                               \\ \cline{2-7} 
                                      & ChainSmith \cite{zhu2018chainsmith} & \CIRCLE  & \Circle                & \textbackslash  & \textbackslash                                 & \textbackslash                                \\ \cline{2-7} 
                                      & TIMiner \cite{zhao2020timiner}                         & \CIRCLE  & \Circle                & \textbackslash  & \textbackslash                                 & \textbackslash                                \\ \hline
Tactic and Technique & ATT\&CK \cite{attck}                         & \Circle       & \Circle                & \textbackslash             & \textbackslash                                 &  \textbackslash                               \\ \cline{2-7} 
                                      & TTPDrill \cite{husari2017ttpdrill}                        & \Circle       & \Circle                & \textbackslash  & \textbackslash                                 & \textbackslash                                \\ \hline
Attack Graph         & EXTRACTOR \cite{satvat2021extractor}                       & \CIRCLE       & \RIGHTcircle                & \RIGHTcircle  & \RIGHTcircle                      & \RIGHTcircle                                \\ \cline{2-7} 
                                      & AttackG \cite{li2021attackg}                        & \CIRCLE       & \RIGHTcircle                & \RIGHTcircle  & \RIGHTcircle                      & \RIGHTcircle                                \\ \cline{2-7} 
                                      & CRUcialG                        & \CIRCLE       & \CIRCLE                & \CIRCLE       & \CIRCLE                           & \CIRCLE                          \\ \hline
\end{tabular}
}
\vspace{-12pt}
\end{table*}

In Table~\ref{table:related-work}, we summarize different attack representations proposed in the existing CTI-based works \cite{liao2016acing,zhu2018chainsmith,zhao2020timiner,attck,husari2017ttpdrill,satvat2021extractor,li2021attackg} that aim to detect APTs. Attack representations can be divided into three categories: attack entity, tactic and technique, and attack graph. We focus on their performance in terms of machine-processable and attack procedure information. The meaning of “machine-processable” is whether the attack representation method can be directly read by a computer. For example, attack entities or attack graphs are structured data that can be read by computers and used for threat search and other tasks, while tactic and technique are too abstract. Besides, regarding EXTRACTOR \cite{satvat2021extractor} and AttackG \cite{li2021attackg}, which are highly relevant to our work, we also compare the performance of their constructed graphs in the aspects of method universality, graph connectivity and graph rationality. Although the attack entity is able to be encoded and recognized by machines, it cannot effectively associate the context information of an attack. Similarly, for tactic and technique, the related methods ignore the procedure information as well (details in Section~\ref{sec:related}). To this end, We will focus on comparing the state-of-the-art (SOTA) methods that extract the attack graph from the CTI reports. The shortcomings of existing research studies are as follows: 1) \textbf{Universality.} It refers to the universality of the ASG reconstruction method. CTI reports come from a wide range of sources, and due to the differences in writing habits, language, and style of security practitioners, the content of reports written by different authors varies greatly. Extracting attack knowledge from CTI reports is key to building graphs. Previous studies utilized security expert experience or syntax dependency analysis techniques to manually \cite{milajerdi2019poirot} or design complex pipelines \cite{satvat2021extractor,li2021attackg} to extract IOC and threat actions from CTI reports. Due to the multi-source heterogeneity of CTI reports, analysts need to observe different styles of language to customize the NLP pipeline, and it cannot cover the varied types of reports. 2) \textbf{Rationality.} It refers to the rationality of the graphs constructed from CTI reports. Existing studies \cite{satvat2021extractor,li2021attackg} lack rational considerations when extracting ASG. CTI reports can display contextual behavioral information of the attacker. However, due to the different understanding of the attack by the authors or the different focus of the description of the attack phase (for example, some operations of accessing sensitive files are abbreviated as “do some information collection” in the report). This results in that the CTI report being unable to provide complete attack information, and the details of the attack cannot be completely reproduced using only traditional information extraction methods. As a result, the reconstructed ASG cannot truly reflect the complete attack scenario. 3) \textbf{Connectivity.} It refers to the connectivity of the graphs constructed from CTI reports (absence of large number of split subgraphs and free nodes). CTI reports contain strong domain expertise, including specific terminologies and abbreviations, with complex grammatical and syntactic structures. More importantly, key entities and relations in CTI reports usually require specific context to be correctly understood, and analysts cannot rely solely on NLP technology for effective adaptation. For example, EXTRACTOR \cite{satvat2021extractor} does not consider the identity transformation of process entities in the context description, and merges non-IOC entities into one entity, resulting in a star-like attack graph (the attack procedure is inconsistent). AttackG \cite{li2021attackg} does not filter redundant text, and the extracted threat actions lack type division and related temporal information. Finally, the attack graph presented by AttackG has problems such as subgraph fragmentation and context information loss. A running example will be detailed in Section~\ref{sec:Problem}.
\par
To fill the key research gaps mentioned above, we propose a universal, rational, and connected ASGs reconstruction system, called \sn\footnote{In CRUcialG, “C” stands for Connectivity, “R” stands for Rationality, and “U” stands for Universality, indicating that the system is able to reconstruct the connected and rational attack scenario graphs using a universal approach.}. \sn{} is able to automatically extract entities and relation from CTI reports to form preliminary ASGs, evaluate the rationality of ASGs from tactical phase with attack procedure, repair and complete ASGs to ensure the connectivity and rationality. Overall, the contributions of this article are as follows:

\begin{itemize}
\item{We build a universal entity/relation extraction module to automatically parse CTI reports. Different from traditional NLP manual design rules and feature engineering techniques that rely on domain knowledge, this study is inspired by large language models and uses the data labeled by security experts to perform transfer learning and training of deep language models, and adaptively realizes the mapping from natural language text to ASGs.}

\item{We propose a phase verification method based on the attacker’s intention. By observing a large number of CTI reports and gaining insights into the entire attack lifecycle, we construct a four-phase attack rationality validation framework by incorporating ATT\&CK, and evaluate the rationality and completeness of the ASG based on attack techniques, tactics and procedures.}

\item{We present a serialized graph generation model, through which we learn the potential attack patterns of the complete attack graph, capture the node features, edge features and time information in the graph, and realize relation repair and phase supplement combined with the four-phase attack rationality validation framework. }

\item{We collect a total of 10,607 CTI reports and generate 5,761 complete ASGs. We conduct a detailed evaluation of all the modules in \sn{}. Experimental results on CTI reports from 30 security vendors and DARPA show that the similarity of ASG reconstruction by \sn{} can reach 84.54\%. Compared with SOTA (EXTRACTOR and AttackG), the recall of \sn{} (extraction of attack events) can reach 88.13\% and 94.46\% respectively, which is 40\% higher than SOTA on average. The F1-score of attack phase verification is able to reach 90.04\%. 
}

\end{itemize}
\par
\textbf{Usage Scenario.} In practice, ASGs reconstructed by \sn{} not only help threat analysts to build threat analysis models, but also provide deeper insights into how attacks are carried out, so as to identify and understand known attack patterns or potential new types of threats, and to take appropriate response countermeasures. In addition, we propose a new attempt for anomaly detection in Section~\ref{sec:discuss}, which combines selected ASGs to better constrain the decision boundaries and reduce false alarms. With the rise of big language model, extracting key threat knowledge from the rich corpus of CTI reports can also help train security big models (e.g., security Q\&A models).
\par

\section{Motivation Example}\label{sec:Problem}

To illustrate the key scientific questions addressed in this work, we provide a detailed comparative discussion with two related studies (EXTRACTOR \cite{satvat2021extractor} and AttackG \cite{li2021attackg}) that are highly relevant to us via a running example.
\par

\begin{figure*}[h!t]
\centering
\includegraphics[width=\textwidth]{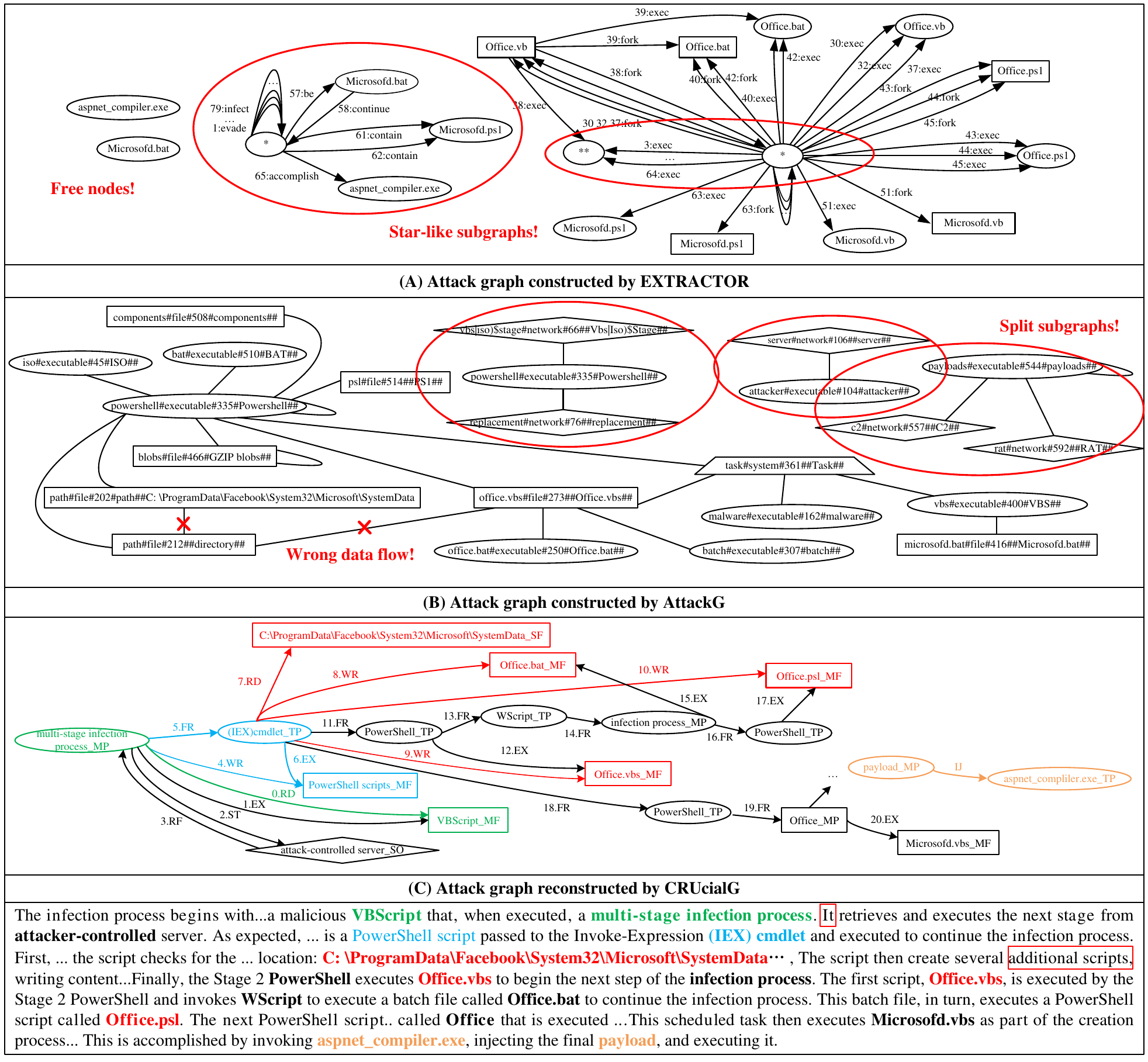}
\setlength{\abovecaptionskip}{-12pt}
\setlength{\belowcaptionskip}{0pt}
\caption{A motivating example. Subfigure (A) and Subfigure (B) are the attack graphs constructed from the AsyncRAT \cite{AsyncRAT} report by EXTRACTOR and AttackG respectively. Subfigure (C) shows the attack scenario graph (omit some nodes and edges) reconstructed by \sn{}. The key events matched in the four attack phases are marked in green, blue, red, and orange, respectively (will be mentioned in Section~\ref{sec:sysdesign_rational}). Below the graph is a brief description of the report, which identifies the key entities extracted. The complete redundancy-filtered text can be found in Appendix~\ref{appendix:AsyncRAT}.}.
\label{fig:graph example}
\vspace{-10pt}
\end{figure*}

Figure~\ref{fig:graph example} shows the attack graph constructed from the CTI report AsyncRAT \cite{AsyncRAT} by EXTRACTOR, AttackG, and \sn{}, respectively. The filtered text for AsyncRAT is shown at the bottom of the figure.  
Subfigure (A) shows the attack graph constructed by EXTRACTOR with customized NLP Pipeline. However, customized NLP pipelines obtain syntactic links and dependencies between words in a sentence mainly through dependency parsing. In particular, system call relations between entities in a sentence are mainly obtained by constructing a verb dictionary.
If the complexity of the sentence increases, dependency parsing is not able to detect tags and inter-word relations. In addition to that, if verbs not mentioned in the dictionary are recognized, there is no way to map the dependencies between entities. More importantly, EXTRACTOR merges non-IOC entities of the same type. As a result, it leads to the construction of an attack graph that fails to show information about the attack procedure of different entities through multiple attack behaviors to achieve attack goals, and loses the contextual behavioral semantics of the attack. As shown in Subfigure (A), not only does the attack graph have split subgraphs, but the subgraphs exhibit a star-like. Overall, the customized NLP pipeline is not sufficiently adapted to multi-source heterogeneous and semantically complex CTI reports, leading to a large number of free nodes and subgraph splits in the final generated attack graph. Subfigure (B) shows the attack graph constructed by AttackG. AttackG proposes an innovation in Named Entity Recognition (NER) model and attack technique template graphs. It combines the extraction of attack graphs with attack techniques and tactics. However, compared to EXTRACTOR, AttackG lacks the filtering step of attacking irrelevant text, which leads to generating a large number of split subgraphs, i.e., it is unable to show a complete and accurate attack scenario. In addition, AttackG lacks the extraction of key dependency actions, and the relation connections between entities lack reasonable data flow (for example, in Subfigure (B) there is a dependency between the file and the file entity), resulting in an attack graph that fails to display effective attack information.
\par

As shown in Subfigure (C), \sn{} subdivides the entity types according to different behavioral purposes from the perspective of attackers, where the definition of entity types is shown in Appendix Table~\ref{table:graph-entity-inter}, and the specific relation types are shown in Appendix Table~\ref{table:graph-relation}. \sn{} builds a universal entity/relation extraction module to automatically parse CTI reports, proposes a phase verification method based on the attacker’s intention, and presents a serialized graph generation model to realize relation repair and phase supplement. Therefore, the comparison of the above examples reveals that the ASG reconstructed by \sn{} in Subfigure (C) is connected and rational (Subfigure (C) will be explained in detail in Section~\ref{sec:sysdesign_rational}). \sn{} easily overcomes the limitations of the previous work as stated above.

\section{System Design}\label{sec:sysdesign}

\subsection{Design Goals}\label{subsec:design goals}
\begin{figure*}[h!t]
\centering
\includegraphics[width=\textwidth]{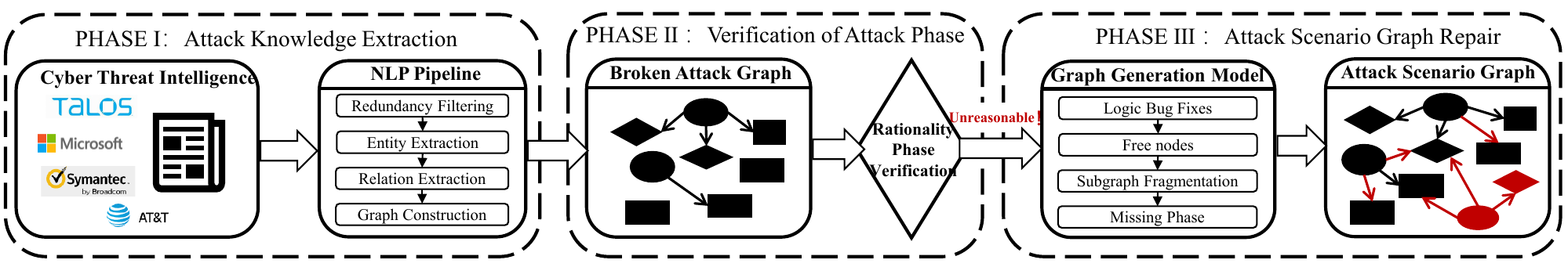}
\caption{The architecture of \sn{}. First, \sn{} extracts attack knowledge and builds a graph from CTI reports. Second, the attack rationality verification framework is used to judge whether the ASG is rational. Finally, the ASG is repaired and supplemented by the graph generation model.}
\label{fig:System Architecture}
\vspace{-12pt}
\end{figure*}

The basic architecture of \sn{} is shown in Figure~\ref{fig:System Architecture}. The design goals of \sn{} are as follows:
\textbf{G1: Attack Knowledge Extraction.} The system should be able to automatically extract attack scenario elements (including entities and the interactions between entities) from CTI reports to build preliminary ASGs. 

\textbf{G2: Attack Rationality Verification.} The system should be able to construct the attack phase verification framework from the attack tactical level to judge the rationality of the preliminary ASGs.
\textbf{G3: Attack Scenario Graph Repair.} The system should be able to perform relation repair and attack phase supplementation by graph generation model for incomplete ASGs.
\par

\subsection{Attack Knowledge Extraction}
\label{sec:sysdesign_nlp}
\par 
In this section, we will explain in detail how to extract attack-related entities and relations from unstructured CTI reports to build ASGs that can be mapped to system-level logs. CTI reports are written in natural language and there is a huge semantic gap in converting this highly complex description form into a graph data structure. It brings great challenges to design an automated extraction pipeline. First, a CTI report consists of several complex sentences. Due to the particularity of human natural language description, these sentences may not fully describe the actual attack content, that is, the report contains a large number of redundant non-attack-related sentences (such as a simple introduction to malware functions, etc.), which may interfere with the extraction of knowledge. Second, each attack-related sentence contains IOC entities and non-IOC entities. IOC entities are domain-specific terms with strong structural characteristics (e.g., file HASH, IP address, registry, etc.). Non-IOC entities are represented by various words in the text. Considering the ubiquitous reference problem in a text description, non-IOC entities can be divided into referents (e.g., “malicious file”, “infection process”, etc.) and pronouns (e.g., “it”, “that”, etc.). Finally, to build an ASG, it is also necessary to extract the interactive actions between entities from the text. Considering the association of attack context, the extraction of relation involves both intra-sentence and inter-sentence. Existing work has tried to solve the above problems, but there are still defects such as relation loss, context entity identity conversion information omission, and subgraph fragmentation. Therefore, we design and implement an automated and universal pipeline to extract entities and relations. In short, our pipeline is divided into four components: 1) redundancy filtering, 2) entity extraction, 3) relation extraction, and 4) graph construction.
\par

\textbf{IOC protection and redundancy filtering.} In order to deal with the large number of redundant texts in CTI reports and maximize the accuracy of extraction, we filter irrelevant sentences at the sentence level by training a de-redundant model. First, we use NLTK \cite{nltk} to segment the text to get the sentence boundaries correctly. However, this tool only considers classic punctuation marks such as “.”, “!”, and “?” when demarcating sentence boundaries. When faced with standard IOC entities such as “[.]” and “. exe” included in CTI reports that may affect NLTK clauses, this will be a problem. To achieve this, we create a set of standard matching rules to identify various types of IOC entities. For example, IP addresses, URLs, and files with different extensions. We then replace the IOC entity with its corresponding attribute words and record its position in the sentence. This ensures accurate extraction of IOC entities without interference from domain-specific symbols, while maintaining proper sentence segmentation. Then, we restore the IOC entities according to the position record. The insight behind this approach is that the attack-related sentences often contain more IOC entities, and to ensure that the de-redundancy model can learn the features of attack-related sentences, the strong domain features of IOC entities are retained. And we believe that redundancy in CTI reports refers to sentences unrelated to the attack that are written by the authors to enrich the content of the report. Finally, since the BERT \cite{devlin2018bert} model has a bidirectional transformer encoder, it will consider the left and right context of a word at the same time. We construct the dataset to train the BERT model to filter attack-irrelevant sentences. The specific model performance will be explained in detail in Section~\ref{sec:evaluation}.
\par

\textbf{Non-IOC entity extraction.} In CTI reports, non-IOC entities account for approximately 60\% of the entities in an article (statistics from experimental data sets). As shown in Appendix Table~\ref{table:graph-entity-inter}, we make a more detailed division of entity types according to the behavior of attackers, and further use them in entity dataset labeling and identification. Unlike the previous works \cite{alam2023looking, ren2022cskg4apt, ahmed2024cyberentrel, zacharis2023aicef, jin2023darkbert, lim2017malwaretextdb, wang2020dnrti}, our aim is to identify the types of entities (i.e., process, file, and socket in Table VIII) involved in the attack procedure and their interactions, in order to construct attack scenario graphs (ASGs) that facilitates downstream security analysis tasks. In the non-IOC entity extraction module, we borrow from the standard span-based NER model \cite{lee2017end, luan2018multi, wadden2019entity, zhong2020frustratingly} in the field of named entity recognition. First, we use a pre-trained language model to obtain the context representation $h_t$ of each token, given a span $s_i \in S$, the beginning and ending token of the target span will be defined as $b_i$, $e_i$, and the representation of this span $\textbf{x}_h(s_i)$ is defined as follows:
\begin{equation}
\scalebox{1.2}{$
\textbf{x}_h(s_i) = [\textbf{h}_{b_i};\textbf{h}_{e_i}; \phi (s_i)]
$}
\vspace{-3pt}
\end{equation}
Where $\phi (s_i)$ represents the width of the span feature embedding. The context representation of the beginning and ending token of each span, as well as $\phi (s_i)$, are spliced together to obtain the representation of the span and then sent to the two-layer feedforward neural network (FFN) to predict the entity type. 
In addition, we also applied the cross sentence contextual information mentioned in PURE \cite{zhong2021frustratingly}. We merge cross sentence contexts by extending sentences to a fixed window size $W$ of entity and relation models. Specifically, given an input sentence with $n$ words, we add $(W-n)/2$ words from the left and right contexts respectively. This mechanism can be used to help predict entity types, especially for pronominal mentions. It is worth mentioning that, in order to reduce the labeling cost, we let the BERT model perform enhanced training of domain text on 5,761 CTI reports to help the model better understand the context of entities. We also evaluated other related NER works in RQ1 to demonstrate the effectiveness of our method in extracting CTI entities.
\par

\textbf{Relation extraction.} CTI reports often refer to the same entity using different descriptions in multiple sentences (including explicit and implicit co-reference), making the identification of co-reference entities and the extraction of cross-sentence relation between entities a challenging task. For example, “It” in Figure~\ref{fig:graph example} is a pronoun, referring to the “multi-stage infection process” in the previous sentence (explicit co-reference), while “additional scripts” in the fifth sentence refers to various malicious files below, such as “Office.vbs” (implicit co-reference). In contrast to previous studies \cite{husari2017ttpdrill,satvat2021extractor,li2021attackg} that rely on building verb dictionaries for relation extraction and calculating node alignment scores for co-reference resolution, we extract different types of relations between entities (as defined relation types in Table~\ref{table:graph-relation}) and address the aforementioned issues simultaneously by utilizing the latest document-level relation extraction model KD-DocRE\footnote{https://github.com/tonytan48/KD-DocRE} \cite{tan2022document} in the field of relation extraction. The advantage of document-level relation extraction model is that both intra-sentence relation and complex cross-sentence relation is well considered and captured. In our work, the relation extraction model (adopted with an enhanced BERT pre-trained language model) is to obtain a context-enhanced entity representation for each entity pair $(e_s, e_o)$, where the context-enhanced representation of the subject $e_s$ (the representation form of the object $e_o$ is the same) is:
\begin{equation}
\scalebox{1.2}{$
z_s = \textbf{tanh}(W_sh_{e_s} + W_cc^{(s, o)})
$}
\vspace{-3pt}
\end{equation}
where $h_{e_s}$ is the representation of entity $e_s$ in document embedding, and $c^{(s, o)}$ represents the context vector between the entity pairs $(e_s, e_o)$, i.e., the enhanced representation of each entity in the entity pair is a fusion of the context vector and the entity representation. After obtaining the enhanced representation of the entity pair of the subject and object, the representation of the entity pair is as follows:
{\setlength\abovedisplayskip{0.1cm}
\setlength\belowdisplayskip{0.1cm}
\begin{equation}
\scalebox{1}{$
\begin{aligned}
&g_i^{(s,o)} = \sum_{k}^{j=1}(z_s^j W_{g_i}^j z_o^j) + b_i \\
&g^{(s,o)} = [g_1^{(s,o)}, g_2^{(s,o)}, ..., g_d^{(s,o)}]\\
\end{aligned}
$}
\end{equation}}
The entity embedding $z_s$ is split into $k$ equal-sized groups, such that $z_s = [z_s^1,z_s^2,..., z_s^k]$. $g_i^{(s,o)}$ is the representation of our entity pair in each dimension. $W_{g_i}^j$ is the weight matrix for dimension $i$. $g^{(s,o)}$ is our final entity pair representation. The entity-to-context enhancement representation is encoded by an axial attention mechanism that is concerned with the one-hop neighbor of the entity two-hop triplet. After obtaining such a triple sequence: $\{(MP, Fork, MP'), (MP', Exec, MF)\}$, the model often focuses on finding whether there is some potential relation between the two-hop nodes $MP$ and $MF$, such as $Write$. This potential relation $(MP, Write, MF)$ can give an explanation of the source of $MF$ (i.e., $MF$ is a malicious executable script created by a malicious process) and enrich the semantic expression of the attack graph. The axial attention representation of the entity pairs is calculated by self-attention along the height axis and width axis, and finally, the category of relation between the entity pairs is predicted using an FFN classifier.
\par
\textbf{Graph construction.} Based on the entities and relation extracted by the above model, we first merge co-reference nodes, then add temporal information to edges through the relative positions of subject and object in the text, and finally generate a preliminary ASG.

\subsection{Attack Rationality Verification}
\label{sec:sysdesign_rational}
\par 
Different from other related work on graph construction \cite{satvat2021extractor,li2021attackg,gao2021enabling}, we additionally consider the rationality of generated ASGs. If the ASGs are reconstructed directly based on the CTI report, the usability of ASGs cannot be guaranteed (i.e., incomplete attack semantics). Therefore, in this section, we will describe how to design a novel framework to verify the rationality of ASGs.
\par

\textbf{Four-phase rationality verification framework.} The main purpose of \sn{} is to generate complete ASGs by CTI reports. This graph not only reflects the attack facts described in the report but also restores the entire attack phase of the attacker. Inspired by previous work (HOLMES \cite{milajerdi2019holmes}, MORSE \cite{hossain2020combating}, APTSHIELD \cite{zhu2023aptshield}, and Conan \cite{xiong2020conan}) and deep insights into APT attack phases, we combine the ATT\&CK to conduct in-depth insights into threat events and find that attackers who want to infiltrate an unknown system and achieve their attack objectives need to complete a series of phased targets (flag events). We collected 1,000 CTI reports with both technical and tactical markings from 27 security vendors, and the statistical results are shown in Figure~\ref{fig:Tactic statistics}. It can be found that security experts prefer to describe intermediate tactics such as code execution, defense evasion, discovery, and privilege escalation in CTI reports. Based on the above findings, we propose a four-phase rationality verification framework abstracted from ATT\&CK tactics:

\begin{figure}[ht]
\centering
\includegraphics[width=\linewidth]{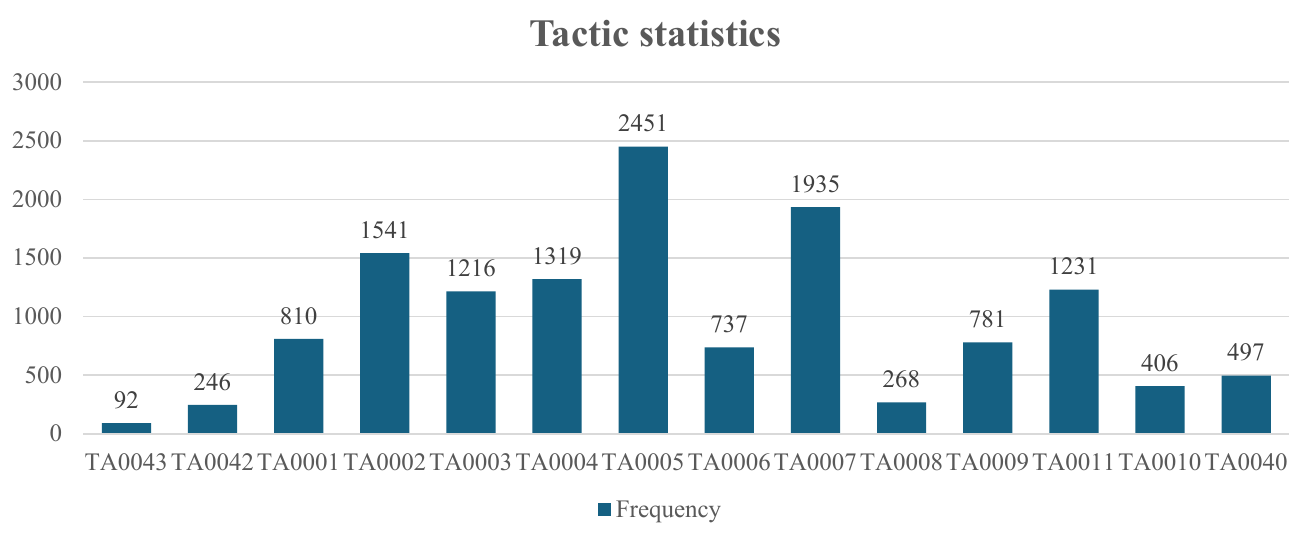}
\centering
\caption{Tactics statistics are derived from the frequency of occurrence of each tactic in 1,000 CTI reports.}
\label{fig:Tactic statistics}
\vspace{-5pt}
\end{figure}

\begin{itemize}
\item{\textbf{Initial intrusion}: This phase corresponds to TA0001 (Initial Access). Initial intrusion consists of techniques (include targeted spear phishing and exploiting weaknesses on public-facing web servers, etc.) that use various entry vectors to gain their initial foothold within a network. The foothold gained through the initial intrusion allows the attacker to have continuous access.}

\item{\textbf{Malicious code/command execution}: This phase includes TA0002 (Execution) and TA0003 (Persistence). In this phase, the attacker will search for vulnerabilities in the target environment’s operating system, application programs, or network services after entering the victim’s network, deliver malicious software, and execute malicious code to obtain long-term operation and latency on the target system.}

\item{\textbf{Break through access control}: This phase includes TA0004 (Privilege Escalation), TA0005 (Defense Evasion), and TA0006 (Credential Access). After attackers successfully execute malicious code, they often take actions to bypass or subvert the system’s access control mechanism to gain unauthorized operations, i.e., attackers try to penetrate the access control layer to further invade the system, obtain sensitive data, expand the scope of the attack.}

\item{\textbf{Leakage and destruction}: This phase includes TA0007 (Discovery), TA0009 (Collection), TA0010 (Exfiltration), TA0011 (Command and Control), and TA0040 (Impact). This phase can be understood as the destructive impact that the attacker implements to achieve his ultimate intention after a series of malicious operations, such as interfering with the operation of normal applications, stealing asset information, collecting and leaking user-sensitive information, and system Deletion and destruction of important data, etc.}
\end{itemize}

\begin{table*}[ht]
\caption{Definition of flag events for attack phase verification. These include key entities and key actions, with restrictions on relevant entities in flag events. The entity type is represented by characters as shown in Appendix Table~\ref{table:graph-entity-inter}.}
\label{table:phase definition}
\resizebox{\linewidth}{!}{%
\begin{tabular}{|c|c|c|c|>{\raggedright\arraybackslash}m{12cm}|} 
\hline
\textbf{Attack Phase}                              & \textbf{Key Entity}                          & \textbf{Key Action}                                               & \textbf{Flag Event}                                    & \textbf{Constraints}                                                                                                                                                                      \\ 
\hline
\textbf{Initial intrusion}                 & \{\textit{MP, TP, MF, SO}\} & \{\textit{Read, Send, Receive}\}                 & Untrusted\_Read(\textit{P, MF})                  & $Time\_forward\{MF_{degree} \gets 0 \}$                                                                                                                                                                \\ 
\cline{4-5}
                                                   &                                              &                                                                   & Malicious\_Link\_Clicks(\textit{P, SO})          & $Time\_forward\{SO_{degree} \gets 0\}$                                                                                                                                                                 \\ 
\hline
\textbf{Malicious code/command execution} & \{\textit{MP, TP, MF}\}     & \{\textit{Fork, Exec, Read, Write}\}             & Target\_Process\_Fork(\textit{MP, TP})                 & $Time\_forward\{Exec \in Interaction(MP, MF)\}\ \&\&\ Time\_backward\{TP_{degree} \neq 0\}$                                                                                                                 \\ 
\cline{4-5}
                                                   &                                              &                                                                   & Malicious\_File\_Write\_Exec(\textit{MP, MF})          & $Time\_forward\{\exists\ Untrusted\_Read(MP, MF')\ \&\ Fork \in Interaction(MP, MP')\}$                                                                                                                       \\ 
\cline{4-5}
                                                   &                                              &                                                                   & Call\_OS\_API(\textit{MP, TP})                         & $Fork \in Interaction(MP, TP)\ \&\&\ Time\_forward\{Exec \in Interaction(MP, MF)\ \&\ TP_{degree} \gets 0\}\ \&\&\ Time\_backward\{TP_{degree} \neq 0\}$                                                 \\ 
\cline{4-5}
                                                   &                                              &                                                                   & Shell\_Exec(\textit{P, MF})                           & $\exists\ Initial\_Intrusion(P)\ \&\&\ Time\_backword\{P_{degree} \neq 0\}$                                                                                                                                    \\ 
\cline{4-5}
                                                   &                                              &                                                                   & Exploitation\_for\_Client\_Execution(\textit{TP, MF})  & $\{Read \in Interaction(TP, MF)\ ||\ Write \in Interaction(TP, MF)\ \&\&\ Time\_forward\{Interaction(TP, SO) \neq \phi\}\}\ \&\&\ Time\_backward\{TP_{degree} \neq 0\}$                                \\ 
\cline{4-5}
                                                   &                                              &                                                                   & Inter-Process\_Communication(\textit{MP, MP'})          & $Time\_forward\{Interaction(MP, SO) \neq \phi\}\ \&\&\ Time\_backward\{MP'_{degree} \neq 0\}$                                                                                                          \\ 
\cline{4-5}
                                                   &                                              &                                                                   & Abuse\_Task\_Scheduling(\textit{MP, TP})               & $Fork \in Interaction(MP, TP)\ \&\&\ Time\_forward\{MP_{degree} \neq 0\}\ \&\&\ Time\_backward\{TP_{degree} \neq 0\}$                                                                                    \\ 
\hline
\textbf{Break through access control}      & \{\textit{MP, TP, MF, SF}\}     & \{\textit{Inject, Chmod, Read, Write}\}          & Abuse\_Elevation(\textit{MP, SF})                      & $Chmod \in Interaction(MP, SF)\ \&\&\ Time\_forward\{MP_{degree} \neq 0\}$                                                                                                                             \\ 
\cline{4-5}
                                                   &                                              &                                                                   & Trigger\_Execution(\textit{MP, TP})                    & $Inject \in Interaction(MP, TP)\ \&\&\ Time\_forward\{Interaction(MP, SF) \neq \phi \}\ \&\&\ Time\_backward\{TP_{degree} \neq 0\}$                                                                    \\ 
\cline{4-5}
                                                   &                                              &                                                                   & Modify\_Access\_Tokens(\textit{P, SF})                & $\{Read, Write\} \subseteq Interaction(P, SF)\ \&\&\ Time\_forward\{P_{degree} \neq 0 \}$                                                                                                            \\ 
\cline{4-5}
                                                   &                                              &                                                                   & Configure\_System\_Settings(\textit{P, SF})           & $\{Read \in Interaction(P, SF)\ ||\ \{Read, Write\} \subseteq Interaction(P, SF)\}\ \&\&\ Time\_forward\{P_{degree} \neq 0\}\ \&\&\ Time\_backward\{Interaction(P, MF) \neq \phi\}$                \\ 
\cline{4-5}
                                                   &                                              &                                                                   & Scripts\_Automatically\_Write\_or\_Exec(\textit{P, MF, SF})      & $Write\ or\ Exec \in Interaction(P, MF)\ \&\&\ Time\_forward\{Write\ or \ Read \in Interaction(P, SF)\}\ \&\&\ Time\_backward\{Interaction(P', MF) \neq \phi\}$                                                                                   \\ 
\cline{4-5}
                                                   &                                              &                                                                   & Modify\_System-level\_Processes(\textit{MP, TP, MF})     & $Inject \in Interaction(MP, TP)\ \&\&\ Time\_forward\{Interaction(MP, SF) \neq \phi \}\ \&\&\ Time\_backward\{MP_{degree} \neq 0\}$                                                                    \\ 
\cline{4-5}
                                                   &                                              &                                                                   & Hijack\_Execution\_Flow(\textit{MP, SF, MF})           & $Time\_forward\{Write \in Interaction(MP, MF)\ \& Interaction(MP, SF) \neq \phi \}\ \&\&\ Time\_backward\{TP_{indegree} \neq 0\}$                                                                      \\ 
\hline
\textbf{Leakage and destruction}           & \{\textit{MP, SF, TF, SO}\} & \{\textit{Read, Unlink, Inject, Send, Receive}\} & Account\_Access\_Removal(\textit{MP, SF})              & $Time\_forward\{\exists\ Modify\_Access\_Tokens(MP, SF')\}\ \&\&\ Time\_Backward\{SF_{degree} \gets 0\}$                                                                                                                                \\ 
\cline{4-5}
                                                   &                                              &                                                                   & Collected\_Sensitive\_Information(\textit{MP, SF, TF}) & $Write \in Interaction(MP, TF)\ \&\&\ Time\_forward\{Read \in Interaction(MP, TF)\}$                                                                                                     \\ 
\cline{4-5}
                                                   &                                              &                                                                   & Service\_Stop\_to\_Avoid\_Detection(\textit{MP, SF})   & $Unlink \in Interaction(MP, SF)\ \&\&\ Time\_forward\{MP_{degree} \gets 0\}$                                                                                                                          \\ 
\cline{4-5}
                                                   &                                              &                                                                   & Clean\_Artifacts(\textit{MP, SF, TF})                 & $Time\_forward\{Read \in Interaction(MP, SF)\ ||\ Unlink \in Interaction(MP, TF)\}\ \&\&\ Time\_backward\{Interaction(MP, MF) \neq \phi\}$                                                        \\ 
\hline
\end{tabular}}
\end{table*}

Among them, the uninvolved TA0043 (Reconnaissance), TA0042 (Resource Development), and TA0008 (Lateral Movement) are not considered within the ASGs we reconstructed, the reason is that our ASGs mainly reflect the attacker’s attack on a single host (victim). Even though, our four-phase rationality verification framework is able to cover 96\% of the techniques and tactics in the ATT\&CK framework.

\textbf{Attack phase verification method based on attacker’s intentions}. The phase verification method is proposed to evaluate whether an ASG can pass the rationality verification, it relies on specific flag events, which maps low-level audit events to the four attack phases. The definition of flag events takes into account the specific actions that attackers may take in order to achieve their intentions during the four phases of the attack. These events can be mapped to specific actions based on the rich semantic information (i.e., the multi-type entities and relations) in the attack graph. For example, interaction between the attacker and sockets, invocation of system processes by the attacker, read and write operations on sensitive files, and creation of temporary high-value files. In detail, we consider the following three factors in attack phase verification: 1) The relation in the ASG is represented as a \textbf{time series} to simulate the evolution procedure of the attack; 2) The relations between entities are formed into \textbf{event streams}; 3) Combining the types of entities and their \textbf{in-degree} and \textbf{out-degree} in the graph structure to complete attack phase verification. Table~\ref{table:phase_defintion} shows some examples of our flag event specifications (see section~\ref{sec:eval_rational} for experimental validation of the completeness of the definition of Table~\ref{table:phase_defintion}). A flag event in Table~\ref{table:phase_defintion} is linked to multiple techniques within a tactic, but not to multiple tactics. \sn{} summarizes techniques of these four tactical phases in ATT\&CK, e.g., to achieve initial access and untrustworthy execution, we focus on flag events such as network data interactions and suspicious files creations. During the attack phase verification, we utilize the timeline of the flag event as a reference and set forward and backward constraints on the relevant entities involved in the flag event (In Section~\ref{sec:sysdesign_repair}, Table~\ref{table:phase_defintion} guided graph generation model to reconstruct complete ASGs).
\par
Take Figure~\ref{fig:graph example} Subfigure (c) as an example, we traverse the action sequence in the graph in chronological order and use different colors to distinguish different attack phases. The first phase (marked in green) is the process’s read an untrusted file (match with Untrusted\_Read event). The key event in the second phase (marked in blue) is that the malicious process calls the target process and executes malicious files (match with Target\_Process\_Fork event). The third phase (marked in red) involves the operation of key system-sensitive files, i.e., multiple malicious files are written in the system folder through the execution of the malicious file in the previous phase to achieve continuous process infection of automatic execution of subsequent malicious scripts (match with Scripts\_Automatically\_Write\_or\_Exec event). The last phase (marked in orange) injects malicious payload into the aspnet\_compiler.exe to damage normal system applications (match with Service\_Stop\_or\_Disrupt event). Each phase is divided in the ASG from the matched flag event action sequence to before the start of the flag event of the next phase.
\par

\subsection{Attack Scenario Graph Repair}
\label{sec:sysdesign_repair}
\par
In this section, we will explain in detail how to solve the problems of \textit{free nodes}, \textit{subgraph fragmentation}, and \textit{missing phases}. Due to the scarcity of manually annotated data sets and the complexity of  CTI reports, it is hard to rely only on the extraction of entity/relation models to form a complete ASG (e.g., the SOTA of the document-level relation extraction models \cite{tan2022document} can only achieve F1-score of 65\% on original general-purpose corpus). Therefore, the goal of this module is to build a graph generation model to learn the true distribution of complete ASGs, and then repair the relation and supplement the phases based on the learned distribution and the guidance of the phase verification model.
\par
\textbf{Logic bug fixes.} The extraction of relations between entities in CTI reports involves complex cross-sentence analysis. In addition, the model's identification of relations may ignore the underlying basic logic of the computer system due to the deviation of the style and order of the text description. Therefore, we design an automatic logical repair method to adjust the ambiguous semantic information due to the loss of time or relation. We mainly adjust the relations between entities by defining some rules, for example, child processes can only make subsequent event calls after being created by the parent process, processes cannot fork multiple times, and other logical rules. 

\par
\textbf{Graph generation model.} Inspired by GraphAF\footnote{https://github.com/DeepGraphLearning/GraphAF} \cite{shi2020graphaf} that generating chemical molecular structures. We use a graph generation model that combines the advantages of autoregressive and flow-based methods, effectively learns the distribution rules of heterogeneous graph structures, and realizes the prediction of heterogeneous nodes and edges serially. It is worth noting that our model is able to add additional domain knowledge for rationality and connectivity checking during its iterative sampling process, allowing it to predict valid nodes and edges (e.g., file and socket do not establish a relation with each other). We let the model sufficiently learn the graph structure, node properties, edge properties, and dependency distributions between nodes from manually labeled Ground Truth. For the graph $G = (V, E, T, A)$, the embedding of node $V_i \in V$ in the subgraph structure is $h_i \in \mathbb{R} ^ {n \times k}$. The free nodes set which needs predict relation is defined as $V_f = \{v_x, v_y, ... , v_z | z < n\}, V_f \in V$. We refer to the function that predicts the type of new node $f$ as $NodeMLP$ and the function that predicts the relation between the node $f$ and its previous $k$ nodes as $EdgeMLP$. The specific formula is as follows:
\begin{equation}
\scalebox{1}{$
g_{f,x} = NodeMLP(sum(h_{f-1}))
$}
\end{equation}
\begin{equation}
\scalebox{0.9}{$
g_{f,i} = EdgeMLP(sum(h_{f-1}), h_{f,f}, h_{f,i}), f-k < i < f
$}
\end{equation}
Where $sum(h_{f-1})$ is the embedding of the entire subgraph of the first $f-1$ nodes, and $h_{f,i} \in \mathbb{R} ^ k$ denotes the embedding of the $i$-th node in the embeddings $h_f$. 

\textbf{Relation repair and phase supplement.} For the \textit{free nodes} and \textit{subgraph fragmentation} problem, we consider both free nodes and part nodes of the split subgraph (prioritize the prediction of relations by selecting nodes with the highest in-degree and out-degree). That is, the node $f$ type predicted by $NodeMLP$ is specified. As for the supplement of phases, we use the rationality verification method to determine the \textit{missing phases} of ASG, so that given the structure of a small part of the subgraph of the existing phase, we can predict the node/edge types at the same time to realize the supplement of the missing phase. It should be noted that under the guidance of phase rationality verification, we believe that prediction will stop as long as the relation repair and phase supplementation meet the phase rationality requirements (Due to the generalization ability of graph generation models, this approach is implemented to prevent the generation of a large number of “false positives” that are conceptually reasonable but significantly deviate from the original CTI report).

\section{Evaluation}\label{sec:evaluation}
In this section, we will evaluate the effectiveness of each component of \sn\ and answer some questions: \textbf{RQ1:} How to evaluate the accuracy of \sn\ in reconstructing ASGs? (Section~\ref{sec:eval_sota}) \textbf{RQ2:} How to evaluate the impact of graph repair technology on the accuracy of ASGs reconstruction? (Section~\ref{sec:eval_repair}) \textbf{RQ3:} How to evaluate the rationality of the attack phase definition? (Section~\ref{sec:eval_rational}) 
\textbf{RQ4:} What is the performance overhead of \sn{}? (Section~\ref{sec:eval_overhead}) 

\subsection{Evaluation Preparation}
\label{sec:eval_prepare}
\par
\textbf{Dataset preparation.} To evaluate \sn, we collected 10,607 public reports from 30 security vendors (including Bitdefender \cite{Bitdefender}, Microsoft \cite{microsoft_blog}, VirusTotal \cite{VirustotalBlog}, Talos \cite{talosBlog}, etc.). We pay more attention to some high-quality textual descriptions, such as some vendors' CTI reports with TTP facts or listing the IOCs involved in the attack process at the end of the report (these features ensure that the CTI reports contain the same system-level information as the DARPA \cite{ET-3, ET-5} and KAIROS \cite{kairosatt} reports, as well as single-host level attacks). In addition to this, these reports were filtered by the number of standard IOCs matched by regular expressions. After filtering, we finally obtained 5,761 high-quality CTI reports. Through keyword analysis, we found that these reports contain 10 unique attack methods such as phishing attack, zero-day attack, watering hole attack, supply chain attacks, spear phishing, trojan horses, backdoor attacks, credential theft, malware implantation, living off the land. In order to answer \textbf{RQ1}, three analysts from famous security vendors with 20+ years of security experience label 5,000 sentences for training de-redundant models and use the brat\footnote{https://github.com/nlplab/brat/releases/tag/v1.3p1} labeling tool to annotate entities and relations (Ground Truth) on 122 carefully selected reports. It is worth mentioning that we select CTI reports based on the distribution of attack types to ensure coverage of 10 types of attacks and minimize biases. The annotation format for the entity model dataset is sciERC \cite{luan2018multi}, and the annotation format for the relational model dataset is DocRED \cite{tan2022revisiting}. Among these 122 reports, 110 are public reports from 11 security vendors, 6 are from the DARPA Transparent Computing \cite{DARPA-TC} in KAIROS \cite{cheng2023kairos} (we believe that the textual description of the attack scenarios in KAIROS \cite{kairosatt} utilizes the ASG reconstruction more than the textual descriptions provided by the DARPA TC \cite{ET-3, ET-5}), and 6 are the reports mentioned in POIROT \cite{milajerdi2019poirot}. In order to answer \textbf{RQ3}, we used open-source TTP extraction tools and TTP facts recorded in the reports to extract attack techniques and tactics from 800 randomly selected public CTI reports. 
\par
\textbf{Experimental parameter selection.} Before the experimental evaluation, we need to explain in advance the key parameters involved in the experiment. The first is the selection of the document sliding window $n$. In order to reduce the difficulty of the model in extracting entity relations in longer texts, we counted the number of cross-sentence relations between each pair of entities in the data set and obtained the candidate parameters as 6, 8, and 10, respectively. Another key parameter is $k$, which is the relation between a node and its previous $k$ nodes in the graph generation model. Use the same method to count the number of entities across the relation between each pair of entities and obtain candidate parameters of 8, 10, and 12, respectively. Through experimental grid search, the model parameters with the best performance were 8 and 10 for $n$ and $k$, respectively. We set the sliding window size to 8 to segment the article, where each window spans the length of a sentence. Additionally, the prediction span of the graph generation model is set to 10, predicting the current node and its preceding 10 nodes along with their relationship types.
\par

\subsection{RQ1: How to evaluate the accuracy of \sn{} in reconstructing ASGs?}
\label{sec:eval_sota}
In order to answer \textbf{RQ1}, we first need to state that \sn{} focuses on using CTI reports to generate effective ASGs that can be mapped to the low-level system logs. In addition, we invited experts to reconstruct 110 complete attack scenarios based on the descriptions of 110 CTI reports. These ASGs (Ground Truth) can all pass our phase rationality verification so that our model can learn complete attack semantics. It should be noted that due to the short description content of 6 DARPA reports (the average sentences of these reports is 6), it is difficult to get the complete ASG from the report alone, so we just constructed a simple graph (incomplete graph) as Ground Truth based on the report description. 
\par

\begin{table}[h!t]
\centering
\caption{Performance of de-redundant model in filtering redundant sentences. These metrics are derived from the classification of attack-relevant and irrelevant sentences in a test set of 1,000 sentences.}
\label{table:BERT model}
\resizebox{\linewidth}{!}{%
\begin{tabular}{|c|c|c|c|} 
\hline
\textbf{Scenario}          & \textbf{Precision} & \textbf{Recall} & \textbf{Accuracy}  \\ 
\hline
\textbf{BERT de-redundant} & 77.99\%            & 86.40\%         & 81.00\%            \\
\hline
\end{tabular}
}
\end{table}

\textbf{Effectiveness of de-redundanct model.} In our system, the main source for reconstructing the ASGs is the relevant description of attack knowledge in CTI reports. As mentioned in Section~\ref{sec:sysdesign_nlp}, CTI reports often contain many redundant descriptions that are not related to attacks. Therefore, we trained a BERT-based two-category language model on 5,000 manually labeled sentences (attack-related: attack irrelevant = 1: 1, train set: valid set: test set = 6: 2: 2). The final performance of the model on the test set is shown in Table~\ref{table:BERT model}. From the table, we can see that compared with precision, the model has a better performance in recall (i.e., most attack-related sentences are retained). In our scenario, we prefer to reduce false negatives (high recall) to maximize the complete description of the attack scenario. Through our analysis, we found that false positive sentences are often additional descriptions of malicious attacks or malware by the writer, such as: “Hannabi Grabber collects survey information from the  infected machine, obtains the geographic location of the system via the IPInfo service,  takes screenshots, and eventually transmits that data to an attacker-controlled Discord server in JSON format.” These sentences, though not indicative of an attack, contain entities that aid in establishing co-reference relationships. Distant entity identity conversion is facilitated through co-reference.
\par
\textbf{Effectiveness of NER model.} Entity extraction is essential for reconstructing the ASGs. In our system, we use both regular expressions as well as NER models for entity extraction in CTI reports. To evaluate the performance of the entity extraction model on redundant filtered text, we use 90 public reports out of 122 reports for model training, 20 public reports, 6 DARPA reports and 6 PORIOT reports for model evaluation (training set: valid set: test set $\approx$ 5: 2: 3, this dataset division is also applied to relational models). Since the descriptions of attacks in DARPA reports are relatively neat, we retained the complete descriptions in the 6 DARPA reports. In the experiment, we compared the relevant methods of LADDER \cite{alam2023looking}, CSKG4APT \cite{ren2022cskg4apt}, and CyberEntRel \cite{ahmed2024cyberentrel} for entity extraction in CTI reports (LADDER\footnote{https://github.com/aiforsec/LADDER} open-source the code, while CSKG4APT and CyberEntRel, do not open-source the code. We reproduce the relevant model architectures as described in the paper). Similar to our work, these tasks enhance or fine-tune pretrained language models based on transformers. In addition, we also compare with the SOTA large language models GPT-3.5 and GPT-4 by prompting ChatGPT \cite{openai}.  
We refer to the official OpenAI prompt engineering document \cite{openai_pro} to guide GPT in entity extraction for CTI reports. Here, we combine prompt optimization methods such as RAG, CoVe, and COT. With equal data feeding, experimental results of NER extraction are shown in Table~\ref{table:ner-model}. It can be seen that \sn{}'s NER model outperforms other related works across the board for entity extraction. We analyze the following reasons: 1) BERT-base is able to show better performance in the downstream task of entity extraction after enhanced training with CTI text. 2) While these model architectures were able to perform well in their respective papers, we found our entity categorization to be more fine-grained compared to these three works (for example, LADDER \cite{alam2023looking} classifies the categories of entities as Malware, Attack patterns, Applications, OS, Organizations, People, Time, Location, etc.). 3) These tasks do not consider the effect of document context on entity categorization and the annotation of these datasets does not differentiate the datasets according to the documents, while the sciERC \cite{luan2018multi} annotation format that we adopted divides them in terms of documents. In summary, our entity categorization task is more granular and the correct categorization of entities relies on cross-sentence analysis. In addition, our NER model also outperforms the GPT. We analyze that small models trained for a specific downstream task outperform large models when the experimental dataset is sparse, as described in \cite{BNBSnor2023, hsieh2023distilling}. The performance of NER models is limited by insufficient training samples and complex texts. To improve the completeness of ASG reconstruction, we introduce a graph repair module (see Section~\ref{sec:eval_repair}) to address the limitations of NLP through graph learning.
\par

\begin{table}[]
\centering
\caption{Performance of NER model and GPT on entity extraction results. We consider entity specific instance names as well as entity types during the evaluation process.}
\label{table:ner-model}
\resizebox{\linewidth}{!}{%
\begin{tabular}{|c|c|c|c|}
\hline
                                       & \textbf{Precision} & \textbf{Recall} & \textbf{F1-score} \\ \hline
\textbf{XLM-RoBERTa-base} \cite{alam2023looking}             & 37.65\%            & 50.21\%         & 43.03\%           \\ \hline
\textbf{RoBERTa-BiGRU-CRF} \cite{ahmed2024cyberentrel}              & 18.10\%            & 14.43\%         & 16.07\%           \\ \hline
\textbf{BERT-BiLSTM-GRU-CRF} \cite{ren2022cskg4apt}        &  12.68\%            & 33.19\%         & 18.35\%                  \\ \hline
\textbf{GPT-3.5}                       & 31.10\%            & 11.80\%         & 17.28\%           \\ \hline
\textbf{GPT-4}                         & 40.00\%            & 26.50\%         & 32.06\%           \\ \hline
\textbf{ctiBERT-base (cross-sentence)} & \textbf{61.51\%}            & \textbf{57.64\%}        & \textbf{59.51\%}           \\ \hline
\end{tabular}%
}
\vspace{-10pt}
\end{table}

\begin{table}[]
\caption{The accuracy of \sn{} reconstructing ASG on the test set (consisting of 20 public reports, 6 DARPA reports, and 6 POIROT reports). The calculation of precision and recall is based on comparing Ground Truth and reconstructed ASGs from the perspective of graph structure. Cosine similarity is used to measure the similarity between the reconstructed ASGs and Ground Truth.}
\label{table:eval_ASG_30}
\centering
\resizebox{\linewidth}{!}{%
\begin{tabular}{|>{\centering\arraybackslash}p{3.5cm}|c|c|c|}
\hline
\textbf{CTI Reports} & \multicolumn{3}{c|}{\textbf{CRUcialG vs. Ground Truth}}                                                     \\ \cline{2-4} 
                                      &\textbf{Precision} & \textbf{Recall}  & \textbf{Graph Similarity} \\ \hline
Biasini\_PoetRAT                      & 39.13\%            & 47.37\%          & 71.16\%                   \\ \hline
Gatlan\_malware                       & 46.34\%            & 90.48\%          & 61.16\%                   \\ \hline
Malhotra\_AridViper                   & 73.91\%            & 52.31\%          & 78.20\%                   \\ \hline
Williams\_Honeypot                    & 49.09\%            & 52.94\%          & 58.99\%                   \\ \hline
Svajcer\_XLLing                   & 57.41\%            & 68.13\%          & 42.56\%                   \\ \hline
McKay APT36                           & 83.33\%            & 62.50\%          & 77.46\%                   \\ \hline
David\_Phishing                       & 71.22\%            & 93.40\%          & 95.75\%                   \\ \hline
anonymous\_RATs                       & 77.78\%            & 63.64\%          & 82.12\%                   \\ \hline
Venere\_ransomware                    & 37.70\%            & 69.70\%          & 52.73\%                   \\ \hline
Carter\_TeslaCrypt                    & 64.58\%            & 70.45\%          & 98.49\%                   \\ \hline
Ventura\_StrongPity                   & 79.45\%            & 74.36\%          & 97.84\%                   \\ \hline
Svajcer\_newLoader                    & 54.17\%            & 75.00\%          & 97.42\%                   \\ \hline
Paul\_Kimsuky                         & 70.00\%            & 87.50\%          & 99.28\%                   \\ \hline
Malhotra\_Emotet                      & 71.43\%            & 83.33\%          & 93.39\%                   \\ \hline
Allievi\_Newexploit                   & 38.78\%            & 70.37\%          & 65.25\%                   \\ \hline
Raghuprasad\_Droppr                   & 82.44\%            & 75.78\%          & 97.38\%                   \\ \hline
An\_newmalware                        & 64.91\%            & 88.10\%          & 93.44\%                   \\ \hline
Windsor\_cryptominers                 & 75.00\%            & 68.48\%          & 81.55\%                   \\ \hline
Chetan\_BitterAPT                     & 67.24\%            & 86.67\%          & 95.62\%                   \\ \hline
Asheer\_RATs                          & 41.03\%            & 82.05\%          & 92.60\%                   \\ \hline
E3-CADETS                             & 57.14\%            & 100.00\%         & 71.03\%                   \\ \hline
E3-ClearScope                         & 64.29\%            & 100.00\%         & 85.92\%                   \\ \hline
E3-THEIA                              & 50.00\%            & 83.33\%          & 91.71\%                   \\ \hline
E5-CADETS                             & 57.14\%            & 84.21\%          & 96.38\%                   \\ \hline
E5-ClearScope                         & 52.63\%            & 76.92\%          & 91.07\%                   \\ \hline
E5-THEIA                              & 66.67\%            & 87.50\%          & 94.91\%                   \\ \hline
Carbanak                              & 40.74\%            & 91.67\%          & 90.49\%                   \\ \hline
DeputyDog                             & 75.00\%            & 100.00\%         & 93.70\%                   \\ \hline
DustySky                              & 48.48\%            & 76.19\%          & 90.95\%                   \\ \hline
HawkEye                               & 48.78\%            & 83.33\%          & 79.05\%                   \\ \hline
njRAT                                 & 41.18\%            & 97.22\%          & 90.82\%                   \\ \hline
Uroburos                              & 73.91\%            & 77.27\%          & 96.92\%                   \\ \hline
\textbf{Average}                      & \textbf{60.03\%}   & \textbf{78.76\%} & \textbf{84.54\%}          \\ \hline
\end{tabular}%
}
\vspace{-15pt}
\end{table}

\par
\textbf{Accuracy of reconstructing ASGs.} In order to evaluate the accuracy of \sn{} for reconstructing ASGs, we constructed artificial ASGs on 122 reports filtered by the de-redundant model. In our experiments, we ensure the same dataset partitioning strategy as for the entity models. Moreover, during the evaluation process, we paid more attention to the effectiveness of the system in extracting entity types, dependency types, and complete graph structures, regardless of the specific instance names of entities. The performance of the final system on the experimental test set (consisting of 20 public reports, 6 DARPA reports, and 6 POIROT reports) are shown in Table~\ref{table:eval_ASG_30}. 
We comprehensively consider precision, recall, and graph embedding similarity \cite{han2021sigl} of ASG compared with manually annotated Ground Truth. In order to obtain more accurate embeddings, we train the Graph LSTM network as the encoder and the MLP as the decoder, thereby obtaining a smaller loss in the embedding calculation of heterogeneous graphs. Finally, the similarity value between ASGs and Ground Truth is measured by calculating the cosine similarity between vectors. From Table~\ref{table:eval_ASG_30}, we can see that the recall and graph similarity of reconstructed ASGs can reach 78.76\% and 84.54\%,  respectively. In addition, we also evaluated and compared with the open-source related work EXTRACTOR\footnote{https://github.com/ksatvat/EXTRACTOR} \cite{satvat2021extractor} and AttackG\footnote{https://github.com/li-zhenyuan/Knowledge-enhanced-Attack-Graph} \cite{li2021attackg} (We reproduce the works EXTRACTOR and AttackG using the provided open-source code), as shown in Table~\ref{table:eval_SOTA_30}. In order to compare these two studies on the same data dimension, we also respectively generalize the types of entity and relation extracted by \sn{}. Only three entities (processes, files, and sockets) are considered in the comparison with EXTRACTOR and AttackG. In addition to this, the edge type is not involved in the evaluation when compared with AttackG. From Table~\ref{table:eval_SOTA_30}, we can see that the average similarity of the reconstructed ASGs can reach 88.36\% compared with EXTRACTOR, and 82.09\% compared with AttackG. Table~\ref{table:eval_ASG_30}’s results are rigorously evaluated from the graph level according to the entity/relation types. Despite models labeling the same entity differently due to various contexts, such as SF/TF or MP/TP, our approach shows a \textbf{7\%} increase in precision and \textbf{10\%} improvement in recall over EXTRACTOR, which doesn't distinguish specific entity types, as shown in Table~\ref{table:eval_SOTA_30}. \sn{} has an average improvement of more than \textbf{40\%} in three comprehensive evaluation indicators (EXTRACTOR and AttackG don't involve model training tasks but rely on customized NLP pipelines for entity-relation extraction, limiting their generalizability to complex datasets), indicating that \sn{} is better at covering real graph structures and is very similar to Ground Truth in vector space. 
\par
One may argue that for some of the CTI reports (e.g., Allievi\_Newexploit, Venere\_ransomware, Biasini\_PoetRAT, njRAT, HawkEye), the precision or recall is not very high. The reasons are as follows: 1) The scenarios described in these reports are mostly spam and vulnerability exploits, and the filtered text descriptions are too abstract (few IOC entities); 2) Since the graph generation model has a generalization function (i.e., the graph generation model learns the complete ASGs during training.), the model generates many redundant edges (not in Ground Truth) during relation repair and phase supplement; 3) In the comparison with EXTRACTOR, the reason why some reports (e.g., njRAT, HawkEye) have higher precision than \sn{} is that EXTRACTOR has made corresponding strategies for the extraction of these reports (where njRAT is an example from the EXTRACTOR paper), such as adding the verbs involved in the reports to the verb dictionary, etc. As a result, EXTRACTOR is able to perform well on these reports, whereas it lacks generalizability to public CTI reports. Although some reports are less precise than our manual ASGs, we are more focused on whether the structure of the ASGs is complete and reasonable. As Section~\ref{sec:discuss}, ASGs can be adapted to different downstream tasks. In addition, in order to verify the universality of \sn{}, we analyzed the style of the 20 reports. According to statistics, these 20 reports were written by 19 different security experts. It can be roughly divided into several attack narrative styles, such as describing a threat operation of the APT organization, using new malware for compromising and exploiting application vulnerabilities. Compared to the SOTA, \sn{} consistently achieves higher recall and graph similarity across various writing and narrative styles. We implement experiments related to ASGs helping anomaly detection models with decision boundaries in Appendix \ref{sec:eval_application}. In this way, we can come to illustrate the effectiveness of the \sn{} reconstruction of ASGs.

\begin{table*}[h!t]
\centering
\setlength{\abovecaptionskip}{0pt}
\setlength{\belowcaptionskip}{-5pt}
\caption{Compare the performance of \sn{} in reconstructing ASGs with different systems on the experimental test set. In the comparison with EXTRACTOR and AttackG, only three entities (process, file, and socket) were considered. Furthermore, when compared with AttackG, edge types do not participate in the evaluation. The calculation of precision and recall is based on comparing Ground Truth and reconstructed ASGs from the perspective of graph structure. Cosine similarity is used to measure the similarity between the constructed ASGs and the Ground Truth (when the similarity is negative, it indicates that the two ASGs are very dissimilar).}
\label{table:eval_SOTA_30}
\resizebox{\linewidth}{!}{%
\begin{tabular}{|>{\centering\arraybackslash}p{4cm}|c|c|c|c|c|c|c|c|c|c|c|c|}
\hline
\textbf{CTI Reports} & \multicolumn{6}{c|}{\textbf{CRUcialG vs. EXTRACTOR}}                                                                                                                                                                              & \multicolumn{6}{c|}{\textbf{CRUcialG vs. AttackG}}                                                                                                                                                                          \\ \cline{2-13} 
                                      & \multicolumn{2}{c|}{\textbf{Precision}}                                          & \multicolumn{2}{c|}{\textbf{Recall}}                                             & \multicolumn{2}{c|}{\textbf{Graph Similarity}}              & \multicolumn{2}{c|}{\textbf{Precision}}                                        & \multicolumn{2}{c|}{\textbf{Recall}}                                           & \multicolumn{2}{c|}{\textbf{Graph Similarity}}            \\ \cline{2-13} 
                                      & \textbf{CRUcialG} & \textbf{EXTRACTOR} & \textbf{CRUcialG} & \textbf{EXTRACTOR} & \textbf{CRUcialG} & \textbf{EXTRACTOR} & \textbf{CRUcialG} & \textbf{AttackG} & \textbf{CRUcialG} & \textbf{AttackG} & \textbf{CRUcialG} & \textbf{AttackG} \\ \hline
Biasini\_PoetRAT                      & 43.48\%           & 25.00\%            & 52.63\%           & 15.79\%            & 94.29\%           & 61.45\%            & 57.69\%           & 30.77\%          & 78.95\%           & 84.21\%          & 91.19\%           & 46.35\%          \\ \hline
Gatlan\_malware                       & 51.22\%           & 29.17\%            & 100.00\%          & 33.33\%            & 87.85\%           & 38.09\%            & 51.22\%           & 50.00\%          & 100.00\%          & 19.05\%          & 80.33\%           & 70.22\%          \\ \hline
Malhotra\_AridViper                   & 82.61\%           & 46.81\%            & 58.46\%           & 33.85\%            & 69.05\%           & 92.17\%            & 97.83\%           & 34.85\%          & 69.23\%           & 35.38\%          & 52.51\%           & 43.26\%          \\ \hline
Williams\_Honeypot                    & 72.73\%           & 50.00\%            & 78.43\%           & 25.49\%            & 61.47\%           & 90.06\%            & 60.71\%           & 37.04\%          & 100.00\%          & 58.82\%          & 37.78\%           & 73.55\%          \\ \hline
Svajcer\_XLLing                   & 77.78\%           & 49.02\%            & 92.31\%           & 27.47\%            & 54.02\%           & 95.57\%            & 77.78\%           & 31.48\%          & 92.31\%           & 37.36\%          & 27.43\%           & -2.30\%          \\ \hline
McKay APT36                           & 83.33\%           & 40.00\%            & 62.50\%           & 25.00\%            & 86.47\%           & 11.56\%            & 100.00\%          & 50.00\%          & 75.00\%           & 16.67\%          & 83.91\%           & 51.85\%          \\ \hline
David\_Phishing                       & 74.82\%           & 40.66\%            & 98.11\%           & 34.91\%            & 78.81\%           & 62.70\%            & 74.82\%           & 36.26\%          & 98.11\%           & 31.13\%          & 64.83\%           & 86.77\%          \\ \hline
anonymous\_RATs                       & 88.89\%           & 36.96\%            & 72.73\%           & 51.52\%            & 89.02\%           & 4.43\%             & 96.30\%           & 62.07\%          & 78.79\%           & 54.55\%          & 90.78\%           & 11.61\%          \\ \hline
Venere\_ransomware                    & 54.10\%           & 61.90\%            & 100.00\%          & 39.39\%            & 92.45\%           & 18.80\%            & 53.23\%           & 27.27\%          & 100.00\%          & 18.18\%          & 87.97\%           & 77.15\%          \\ \hline
Carter\_TeslaCrypt                    & 80.21\%           & 58.62\%            & 87.50\%           & 38.64\%            & 91.37\%           & 66.81\%            & 83.00\%           & 36.07\%          & 94.32\%           & 25.00\%          & 88.53\%           & -32.60\%         \\ \hline
Ventura\_StrongPity                   & 86.30\%           & 45.19\%            & 80.77\%           & 60.26\%            & 99.12\%           & 94.54\%            & 95.89\%           & 42.86\%          & 89.74\%           & 50.00\%          & 98.68\%           & -5.88\%          \\ \hline
Svajcer\_newLoader                    & 65.28\%           & 27.55\%            & 90.38\%           & 51.92\%            & 98.06\%           & 64.88\%            & 68.49\%           & 46.55\%          & 96.15\%           & 51.92\%          & 98.73\%           & 10.87\%          \\ \hline
Paul\_Kimsuky                         & 76.00\%           & 39.13\%            & 95.00\%           & 45.00\%            & 99.33\%           & 95.72\%            & 76.00\%           & 50.00\%          & 95.00\%           & 27.50\%          & 99.22\%           & -4.11\%          \\ \hline
Malhotra\_Emotet                      & 73.81\%           & 45.45\%            & 86.11\%           & 41.67\%            & 94.71\%           & 92.90\%            & 73.81\%           & 64.29\%          & 86.11\%           & 25.00\%          & 93.47\%           & 56.26\%          \\ \hline
Allievi\_Newexploit                   & 48.98\%           & 71.43\%            & 88.89\%           & 37.04\%            & 75.86\%           & 98.59\%            & 50.00\%           & 32.00\%          & 92.59\%           & 88.89\%          & 54.72\%           & 21.56\%          \\ \hline
Raghuprasad\_Droppr                   & 91.71\%           & 42.42\%            & 84.30\%           & 12.56\%            & 97.65\%           & 50.59\%            & 100.00\%          & 39.23\%          & 91.93\%           & 22.87\%          & 95.42\%           & 69.40\%          \\ \hline
An\_newmalware                        & 68.42\%           & 70.00\%            & 92.86\%           & 33.33\%            & 93.15\%           & 89.68\%            & 71.93\%           & 33.33\%          & 97.62\%           & 21.43\%          & 92.79\%           & 30.14\%          \\ \hline
Windsor\_cryptominers                 & 78.57\%           & 27.91\%            & 71.74\%           & 39.13\%            & 83.64\%           & 55.26\%            & 95.24\%           & 34.91\%          & 86.96\%           & 40.22\%          & 68.84\%           & -6.02\%          \\ \hline
Chetan\_BitterAPT                     & 75.86\%           & 42.50\%            & 97.78\%           & 37.78\%            & 97.90\%           & 59.80\%            & 77.59\%           & 23.53\%          & 100.00\%          & 8.89\%           & 96.15\%           & 59.67\%          \\ \hline
Asheer\_RATs                          & 50.00\%           & 61.54\%            & 100.00\%          & 41.03\%            & 90.67\%           & 97.20\%            & 47.56\%           & 41.18\%          & 100.00\%          & 17.95\%          & 83.35\%           & 9.21\%           \\ \hline
E3-CADETS                             & 57.14\%           & 30.77\%            & 100.00\%          & 25.00\%            & 75.87\%           & 49.70\%            & 57.14\%           & 50.00\%          & 100.00\%          & 18.75\%          & 58.42\%           & 62.87\%          \\ \hline
E3-ClearScope                         & 64.29\%           & 0.00\%             & 100.00\%          & 0.00\%             & 93.05\%           & 31.48\%            & 64.29\%           & 57.14\%          & 100.00\%          & 44.44\%          & 91.00\%           & 50.19\%          \\ \hline
E3-THEIA                              & 55.00\%           & 13.04\%            & 91.67\%           & 50.00\%            & 99.27\%           & 44.44\%            & 54.55\%           & 53.85\%          & 100.00\%          & 58.33\%          & 98.45\%           & 87.85\%          \\ \hline
E5-CADETS                             & 60.71\%           & 62.50\%            & 89.47\%           & 26.32\%            & 93.32\%           & 67.17\%            & 67.86\%           & 44.44\%          & 100.00\%          & 21.05\%          & 93.58\%           & 42.28\%          \\ \hline
E5-ClearScope                         & 63.16\%           & 50.00\%            & 92.31\%           & 53.85\%            & 93.58\%           & 60.78\%            & 68.42\%           & 57.14\%          & 100.00\%          & 30.77\%          & 94.13\%           & 80.35\%          \\ \hline
E5-THEIA                              & 66.67\%           & 0.00\%             & 87.50\%           & 0.00\%             & 96.85\%           & -54.88\%           & 76.19\%           & 25.00\%          & 100.00\%          & 6.25\%           & 97.20\%           & 68.39\%          \\ \hline
Carbanak                              & 44.44\%           & 31.82\%            & 100.00\%          & 58.33\%            & 89.56\%           & 94.76\%            & 44.44\%           & 32.14\%          & 100.00\%          & 37.50\%          & 83.93\%           & 88.36\%          \\ \hline
DeputyDog                             & 75.00\%           & 42.86\%            & 100.00\%          & 33.33\%            & 90.88\%           & 84.51\%            & 75.00\%           & 10.00\%          & 100.00\%          & 11.11\%          & 82.64\%           & 43.58\%          \\ \hline
DustySky                              & 60.61\%           & 50.00\%            & 95.24\%           & 28.57\%            & 95.50\%           & 78.06\%            & 63.64\%           & 33.33\%          & 100.00\%          & 19.05\%          & 93.99\%           & 24.10\%          \\ \hline
HawkEye                               & 53.66\%           & 56.00\%            & 91.67\%           & 58.33\%            & 68.77\%           & -3.55\%            & 58.54\%           & 20.00\%          & 100.00\%          & 8.33\%           & 50.52\%           & -4.29\%          \\ \hline
njRAT                                 & 42.35\%           & 83.33\%            & 100.00\%          & 13.89\%            & 97.57\%           & -48.28\%           & 42.35\%           & 40.63\%          & 100.00\%          & 36.11\%          & 97.99\%           & 80.92\%          \\ \hline
Uroburos                              & 78.26\%           & 26.32\%            & 81.82\%           & 22.73\%            & 98.38\%           & 38.39\%            & 95.65\%           & 33.33\%          & 100.00\%          & 18.18\%          & 98.33\%           & 96.19\%          \\ \hline
\textbf{Average}                      & \textbf{67.04\%}  & 42.43\%            & \textbf{88.13\%}  & 34.23\%            & \textbf{88.36\%}  & 55.73\%            & \textbf{71.16\%}  & 39.40\%          & \textbf{94.46\%}  & 32.65\%          & \textbf{82.09\%}  & 43.37\%          \\ \hline
\end{tabular}%
}
\end{table*}

\subsection{RQ2: How to evaluate the impact of graph repair technology on the accuracy of ASGs reconstruction?}
\label{sec:eval_repair}
\par
 
In order to answer \textbf{RQ2}, we conducted ablation experiments on the graph generation model on the test set. As shown in Table~\ref{table:eval_NLP}, the 2nd column shows the knowledge extraction results by the entity/relation extraction model and the 3rd column shows the results when we added relation repair and phase supplement. The 5th and 6th rows are our statistics on the number of free nodes and split subgraphs in the graph. The 7th row is the number of ASGs that have passed the rationality verification. It can be found that the recall of ASG without relation and phase completion is low, and graphs contain more free nodes and split subgraphs. After the graph is repaired by the graph generation model, the number of free nodes and split subgraphs of most graphs has been reduced, the recall rate has increased greatly and all ASG phases have been completed, while the precision has decreased. From the analysis of experimental results, the NLP model cannot guarantee the connectivity and rationality of the extracted ASGs, while the graph generation model is able to learn the behavioral semantics of nodes and edges in complex graph structures and predict nodes and edges that are consistent with the attack phase. Due to the generalization function of the graph generation model, more edges are generated during graph repair, which leads to an increase in the false positive rate (however, for Ground Truth, these newly generated nodes and edges are false positives, whereas for the graph structure, they are rational and representative of the attack). We argue that while the model may have generated information that was never extracted as relevant, we do not believe that this information is incorrect. For example, if the ASG is missing the third phase of the attack, the model will make predictions of the next substructures based on the input information of the first two attack phases. The graph generation model with generalization capabilities may do some remaining additional operations (similar to reading some irrelevant system files, deriving other processes that resemble the activities of legitimate processes, which can be interpreted as the attacker taking a long detour before reaching the final goal) before reaching the critical events of the third phase. In summary, compared with the precision of the graph, we pay more attention to the completeness and rationality of the ASGs. Therefore, it is acceptable to sacrifice precision to ensure recall and rationality results.

\begin{table}
\setlength{\abovecaptionskip}{0pt}
\setlength{\belowcaptionskip}{-10pt}
\caption{Performance of NLP model and \sn{} in reconstructing ASGs on the test set.}
\label{table:eval_NLP}
\centering
\resizebox{\linewidth}{!}{%
\begin{tabular}{|>{\centering\arraybackslash}m{4cm}|c|>{\centering\arraybackslash}p{2.5cm}|} 
\hline
                                                            & \textbf{NLP} & \textbf{CRUcialG}  \\ 
\hline
\textbf{Precision}                                           & 72.07\%      & 60.03\%            \\ 
\hline
\textbf{Recall}                                             & 56.73\%      & \textbf{78.76\%}   \\ 
\hline
\textbf{Graph similarity}                                   & 78.34\%      & \textbf{80.76\%}            \\ 
\hline
\textbf{Free nodes}                                          & 12            & 6                  \\ 
\hline
\textbf{Split subgraphs}                                     & 7            & 4                  \\ 
\hline
\textbf{Number of phase-complete ASGs} & 7            & 26                 \\
\hline
\end{tabular}
}
\vspace{-15pt}
\end{table}

\begin{figure}[ht]
\centering
\includegraphics[width=\linewidth]{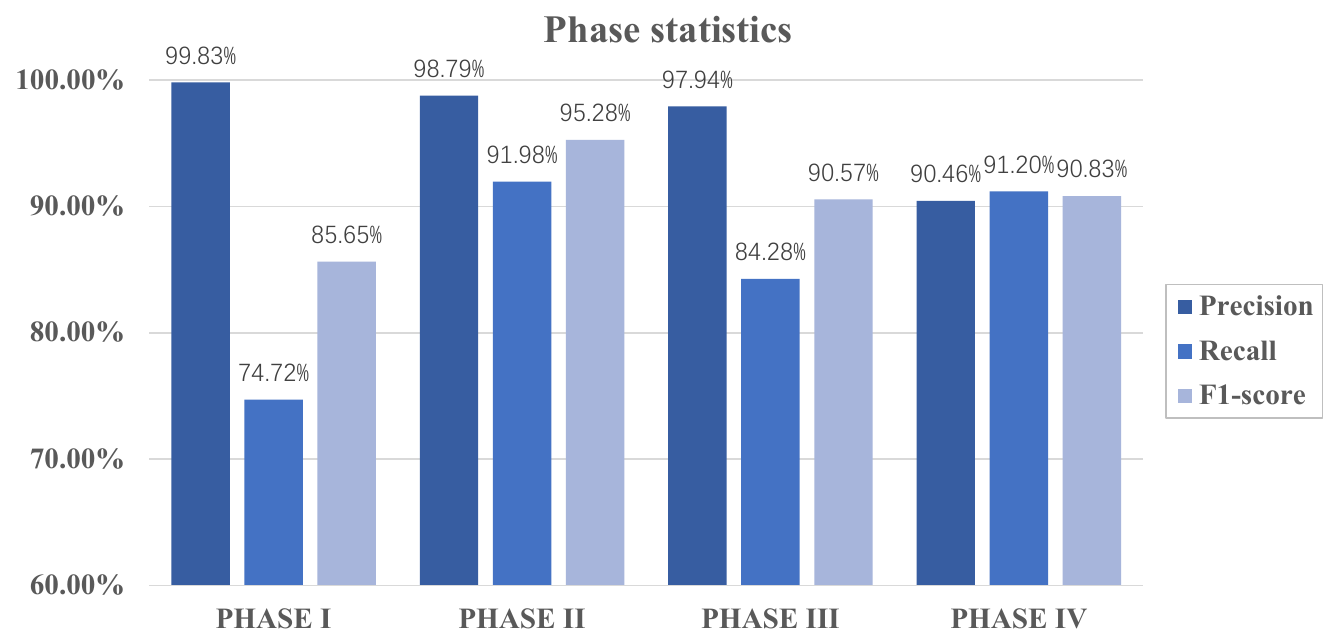}
\centering
\caption{Precision, recall and F1-score of the attack four-phase covered by ASGs which are reconstructed from 800 open CTI reports.}
\label{fig:eval3}
\vspace{-10pt}
\end{figure}

\subsection{RQ3: How to evaluate the rationality of the attack phase definition?}
\label{sec:eval_rational}
For \textbf{RQ3}, we randomly selected 800 public CTI reports with TTP facts recorded from more than 20 security vendors. By adopting several open-source tools (TTDrill\footnote{https://github.com/KaiLiu-Leo/TTPDrill-0.5} \cite{husari2017ttpdrill}, rcATT\footnote{https://github.com/vlegoy/rcatt} \cite{legoy2019retrieving} and TRAM\footnote{https://github.com/center-for-threat-informed-defense/tram}), we obtained attack techniques and tactics described in each report. Further, we mapped these tactics to our four-phase attack procedure. To ensure the accuracy of experimental results, during the process of repairing ASGs, we deleted the guidance of the attack phase verification method in the relation repair process. On the one hand, this experiment can verify the accuracy of ASG reconstruction for phase coverage. On the other hand, it is able to verify the effectiveness of our phase verification definition. After using the graph generation model to repair the ASGs, we use the phase verification method to judge the phase through which the ASGs have passed. The results of the final experiment is shown in Figure~\ref{fig:eval3}. From the result, we can see that the reconstruction of the ASGs has a high phase coverage accuracy on all four attack phases. The only thing that is not so good is the recall of phase I. Through report analysis, we discovered fewer phase I descriptions, with most authors concentrating on detailing the second and third phases of the attack. We respectively calculated the precision, recall and F1-score for a total of four phases and is able to achieve 95.62\%, 85.07\%, and 90.04\%. This also shows that using the graph generation model to repair ASGs is focused on completing the technical and tactical aspects of an attack. 

\subsection{RQ4: What is the performance overhead of \sn{}?}
\label{sec:eval_overhead}
\par 
In order to evaluate the overhead of \sn{}, we calculated the average time and CPU utilization 
of ASG extraction on the test set. We set up the experiment on a host with Intel i9-13900K CPU (24 cores and 3.00 GHz), 64G memory, and GPU (NVIDIA GeForce RTX 4090). As shown in Table~\ref{table:overhead}, the average time for \sn{} to reconstruct an ASG is between EXTRACTOR and AttackG. After analysis, we found that the overhead of EXTRACTOR is mainly concentrated in sentence dependency analysis, while our system overhead is mainly caused by three deep learning models. Among them, AttackG takes a relatively short time because it lacks the classification of relationships. Comprehensive analysis shows that the overhead of all three systems falls within an acceptable range. \sn{} maintains effective graph reconstruction with a suitable overhead, resulting in excellent overall performance.

\begin{table}[]
\centering
\setlength{\abovecaptionskip}{0pt}
\setlength{\belowcaptionskip}{-10pt}
\caption{\sn{} average overhead of reconstructing an ASG on the test set (from the inputs to the outputs of the system).}
\label{table:overhead}
\resizebox{\linewidth}{!}{%
\begin{tabular}{|c|c|c|c|} 
\hline
                    & \textbf{CRUcialG} & \textbf{Extractor} & \textbf{AttackG}  \\ 
\hline
\textbf{Time}       & \textbf{11.71s}   & 154.62s            & 3.55s             \\ 
\hline
\textbf{CPU utilization} & \textbf{0.50\%}   & 0.30\%             & 0.40\%            \\
\hline
\end{tabular}
}
\vspace{-10pt}
\end{table}

\section{Discussion}\label{sec:discuss}
\subsection{Limitation}
\textbf{Attack knowledge extraction.} As shown in Table~\ref{table:eval_NLP}, we can find the problems in which there are many free nodes and split subgraphs after NLP extraction. After analysis, the reasons are as follows: 1) Compared with other studies \cite{satvat2021extractor,li2021attackg,gao2021enabling}, we make a more detailed division of entity and relation types which increases the difficulty of accurate classification of models; 2) Since manually labeling data sets is time-consuming and labor-intensive, the scarcity of the datasets may reduce the effectiveness of multi-classification tasks; 3) The metric of processing co-reference underperformed expectations. Although we use the cross-sentence model to resolve co-references while extracting relations, due to the characteristics of co-reference relation in texts being relatively hidden, the classification effect of co-reference relation in the relation model may not be  good, which in turn leads to more free nodes and split subgraphs. Therefore, in our future work, we can consider an automated dataset annotation method to expand the dataset (for example, consider the use of large language models to aid manual annotation). In the terms of long cross-sentence relations, we can consider additional processing of co-reference relation separately to improve the accuracy of relation classification.

\par
\textbf{Graph generation model.} From the ablation experiment, we can see that although the graph generation model can effectively reduce the number of free nodes and split subgraphs, the precision of the graph decreases after the graph generation model repairs ASGs. This is due to the certain generalization of the graph generation model. In our scenario, we pay more attention to whether the reconstructed ASGs are complete, i.e., through the experiments of RQ3, it can be proved that our graph generation model can ensure that the repaired ASGs satisfy the phase coverage, and can effectively fill in the lack of tactics. The graph generation model is supplemented with more redundant nodes and redundant edges relative to Ground Truth in order to reconstruct the complete ASGs on the results of the NLP extraction. Yet these ASGs themselves fulfill the requirements for a complete rational attack graph.

\subsection{Future Work} 
\textbf{Threat hunting.} Threat hunting aims to uncover attacks that have circumvented traditional security defenses and APT activities that may not have triggered alarms yet. Recall that ASGs reconstructed from the CTI reports include threat behavior patterns and procedures, achieving a true mapping of attack facts from natural language to low-level system logs. Moreover, we combine generalized types of entity/relation and graph generation model to greatly improve threat hunting’s ability to detect unknown attacks. For instance, the query graph constructed by correlating CTI reports in the POIROT \cite{milajerdi2019poirot} can be used to search for threats at the system level. ProvG-Searcher \cite{altinisik2023provg} uses sequence embedding to further simplify subgraph matching to a comparison between the graph representation of the query graph and the provenance graph. The difference is that the query graph is manually constructed by experts based on DARPA reports, while \sn\ automatically reconstructs ASGs from CTI reports to reduce manual efforts. In addition, the ASG we reconstructed is able to maintain the generalizability, integrity, and rationality of the attack phase from the tactical/technical level.

\par
\textbf{Attack classification model on a specific business endpoint.} One of the biggest challenges in anomalous threat detection is how to cover as many normal behavioral patterns as possible. Current research studies (e.g., UNICORN \cite{han2020unicorn}, DeepLog \cite{du2017deeplog}, and other anomaly-based attack detection systems \cite{hassan2019nodoze,liu2018towards,fang2022back,yang2023prographer, cheng2023kairos, rehman2024flash}) learn the normal baseline behavior of the system and look for outliers to discover potential threats. However, the experimental environment of the above studies is relatively limited, and there are problems such as high overhead and high false positives in cross-scenario and cross-host situations. Since the advantage of ASG is that it is easy to be automatically generated and understood by machines, we discuss a new anomaly detection attempt that is expected to be used in our future work. The core point is that since it is difficult to collect a large number of benign behavioral baselines in the wild, can we decouple the problem and deploy anomaly detection models separately on the specific business host? Since there are not many normal businesses on a single host, we are able to build a migration model that uses local data alone on each endpoint for anomaly detection. When building the model, we combine it with the selected ASGs we reconstructed automatically that represent anomalies to better constrain the decision boundary and reduce false positives.
\section{Related Work}\label{sec:related}

\noindent \textbf{Attack knowledge extraction.}
CTI comes from a wide range of sources, and different sources determine the composition of intelligence information to be different. Structured CTI such as OpenIOC \cite{OpenIOC}, STIX 2.0 \cite{STIX}, CybOX\cite{CybOX}, and TAXII\cite{TAXII} are standardized and presented in clear formats and data patterns, usually including IOC such as known malicious file hash, attacker IP addresses, etc. However, data obtained from structured CTI are single-point independent and lack interactive information between IOCs. 
For example, the abstract nature of entity-relation in STIX 2.0 lacks detailed attack procedure information, making it difficult to map them to the provenance-level (e.g., relations in STIX 2.0. “attack-pattern” include: “target” and “use”, while in the provenance-level we use “read”, “write”, etc.). Alternatively, unstructured CTI reports composed in natural language provide contextual information and timelines of attack events. These reports assist analysts in pinpointing the initial intrusion point, the propagation path, and thereby gaining a deeper understanding of the complete attack procedure. There have been many research studies on analyzing CTI and extracting attack-related knowledge, which can be divided into three types: extracting IOC, extracting Tactic and Technique, and constructing graphs: 1) \textbf{IOC extraction.} iACE \cite{liao2016acing} uses regular expressions and the NER model to extract IOC from the report. ChainSmith \cite{zhu2018chainsmith} classifies the extracted IOC into the corresponding activity phase. TIMiner \cite{zhao2020timiner} classifies the field of CTI reports through the extraction of IOC. The above three works only extract IOC entities but do not involve the classification of relation between entities. 2) \textbf{Tactic and technique extraction.} TTPDrill \cite{husari2017ttpdrill}, ActionMiner \cite{husari2018using}, and EX-Action \cite{zhang2021ex} use dependency parsers to extract threat actions from reports and map them to ATT\&CK via attack techniques and tactics. RcATT \cite{legoy2019retrieving} and other studies \cite{yu2022tactics, liu2022threat} also have achieved the same goal with the help of machine learning models. However, these studies have complex process and the recognition of behaviors lacks interpretability.
3) \textbf{Graph construction.} EXTRACTOR \cite{satvat2021extractor} and ThreatRaptor \cite{gao2021enabling} construct graphs from CTI reports through customized NLP pipelines. However, the above studies need to perform part-of-speech dependency analysis through simplification of sentences and construction of verb dictionaries. It not only involves complex NLP domain knowledge but also reduces the extraction performance of complex cross-sentence relations, and extracted graphs cannot fully reflect the information of attack context. AttackG \cite{li2021attackg} proposes the concept of a technical knowledge graph. However, the constructed graph lacks distinction of relationship types, and because the text is not filtered by deduplication operations, the generated graph has a lot of redundancy. Different from previous work, \sn{} uses language models and graph generation models to generate ASGs with node and dependency attributes. In addition, \sn{} also considers the connectivity and rationality of the graph.

\noindent \textbf{Division of attack phases.}
Currently, the continuously evolving network threats have brought about various complex attack scenarios. Dividing attack phases by observing the attack life cycle can help security practitioners better understand, detect, and respond to threats. Diamond Model \cite{caltagirone2013diamond} describes network intrusion events through four core features, including adversaries and victims, and represents the basic relation of different features based on edge connections. Cyber Kill Chain \cite{hutchins2011intelligence} proposes a 7-phase APT intrusion detection model from the perspective of the defender and maps the action process of detection, defense, and response for each phase of the intrusion. The MITRE ATT\&CK \cite{attck} framework deploys tactics and techniques from the attacker. HOLMES \cite{milajerdi2019holmes} customizes a 7-phase APT attack life cycle model and realizes the one-to-one mapping of low-level system audit logs to advanced APT attack phases by building an intermediate layer based on the ATT\&CK model. In order to detect unknown APT attacks and find invariants in the attack process, Conan \cite{xiong2020conan} proposes a syllogism in the attack phase. However, the above-mentioned methods of dividing attack phases focus more on the attacker’s tactics and techniques, and do not describe in detail the potential attack procedure of a real and complete attack. Unlike previous methods, \sn{} proposes a threat verification method based on the attacker’s intentions. Starting from the attacker’s attack intentions at each phase, explore the essence and procedure of the attack, and divide the attack into four phases from the perspective of attack tactics to verify the reasonable effectiveness of ASGs.

\section{Conclusion}\label{sec:conclusion}
In this paper, we propose a system called \sn{} that can automatically reconstruct ASGs by CTI reports. \sn{} is able to extract attack knowledge from CTI reports described in natural language to form preliminary ASGs, evaluate the rationality of ASGs from tactical phase with attack procedure, repair and complete ASGs to ensure the connectivity and rationality. Experimental results demonstrate that \sn{} outperforms the SOTA by an average of 40\% across various metrics.

\bibliographystyle{IEEEtran}
\bibliography{main}

\clearpage
\appendix
\setcounter{equation}{0}
\pagestyle{empty}
\renewcommand{\thesection}{Appendix \arabic{section}}
\setcounter{section}{0}

\section{Appendix}\label{sec:appendix}
\subsection{Background Knowledge}\label{sec:appendix_background}
\subsubsection{Provenance Graph}
Provenance graph is a data structure that describes the origin and evolutionary history of system behavior. By recording the process of data generation, transformation, and transmission, it can track the trajectory of data and provide more contextual evidence to answer various historical questions about the described data. A provenance graph is a directed temporal heterogeneous graph that includes the attributes and temporal information of vertices and edges. It consists of nodes representing subjects (e.g., processes) and objects (e.g., files or sockets), along with the relation between them. As a result, the provenance graph is defined as $G = (V, E, T, A)$, the nodes are defined as $V = \{v_1, v_2, ... , v_n\}$, the directed edges of the set are defined as $E = \{(v_i v_j) | v_i, v_j \in V\}$, temporal set is defined as $T = \{T (e_1), T (e_2),... , t(e_m)\}$, the heterogeneous attribute set is defined as $A = \{a(v_1),..., a(v_n), a(e_1),..., a (e_m) | n + m > 2\}$. Our proposed ASG is a provenance graph built at the host level (i.e., it is a kind of system-level provenance graph). 

\subsubsection{Cyber Threat Intelligence Report}
CTI reports are text information about cyber threats, attackers, attack methods, vulnerabilities, and malwares that are recorded by security practitioners to help organizations predict, defend against, and respond to cyber threats. CTI reports have the following characteristics: 1) \textbf{Multi-source heterogeneous.} CTI reports are generally natural language texts that provide rich attack context information. Different analysts from different security vendors have very different writing styles. 2) \textbf{Strong domain characteristics.} CTI reports are evidence-based knowledge that provides IOCs and factual data about attack activities. It not only covers the characteristics of known threats but also reveals emerging threats and potential behavior patterns of attackers. 3) \textbf{Huge quantity.} With the continuous development of cyber technology, computer security problems also emerge endlessly, and the number of CTI reports has also grown explosively, which has led to the time-consuming and labor-intensive manual analysis of CTI reports. Therefore, we use unstructured CTI reports to automatically reconstruct attacks and obtain ASG to improve the understanding and utilization of cyber attacks.
\par

\subsubsection{Natural Language Processing}
NLP is a crucial subfield of artificial intelligence. From the early usage of rules and grammar knowledge to process text, to the remarkable performance of deep learning models (e.g., Word2Vec \cite{mikolov2013efficient}, GloVe \cite{pennington2014glove}, etc.) on various NLP tasks, NLP research has gradually expanded to multilingual and cross-lingual fields. Recently, pre-trained language models have been developed rapidly, such as Bidirectional Encoder Representations from Transformers (BERT) \cite{devlin2018bert} and Generative Pre-trained Transformers (GPT) \cite{radford2018improving}, which are pre-trained with large-scale data and then fine-tuned to adapt to specific downstream tasks, resulting in stunning effect. However, to extract systematic knowledge from CTI reports with strong domain characteristics, a professional text-processing pipeline should be customized. As we mentioned in Section~\ref{sec:intro}, designing such a text processing pipeline requires a certain threshold of professional knowledge. At present, some research studies use customized dictionaries to simplify the vocabulary or complex sentences in the professional field to assist in extracting entities and relations, which lack generalization and inevitably cause certain information loss. 

\subsubsection{Pre-trained Language Models}
Pre-trained language models are deep learning models that learn general natural language understanding and expressiveness by pre-training on large-scale text corpora (e.g., Wikipedia, news articles, etc.). In the pre-training phase, the model performs a self-supervised learning task to capture the intrinsic structural and semantic information of the language on the text dataset without manual annotation. Currently, pre-trained models are mainly used to fine-tune for downstream tasks, i.e., by fine-tuning the last layer(s) of the model, supervised learning is performed on the training data to adapt the model to the needs of a specific task. One of the most representative is BERT \cite{devlin2018bert}, proposed by Google in 2018, which is a deep learning model based on Transformer’s self-attention mechanism. The pre-training process consists of two main tasks: Masked Language Model (MLM), which randomly masks some words in the input text and then tries to predict these masked words so that the model can understand the semantics of the words in the context. Next Sentence Prediction (NSP), which learns the relationship between text segments by determining whether two sentences are adjacent in the corpus, helps the model understand the relevance between texts. With these two tasks, BERT is able to consider both the left and right contexts of a word to better understand contextual information.

\begin{table*}[]
\caption{The attribute of system entities in \sn{}. The second column are types of entities and corresponding abbreviations. The third column is an explanation of the entity type definition.}
\centering
\label{table:graph-entity-inter}
\resizebox{\linewidth}{!}{%
\begin{tabular}{|c|c|c|}
\hline
\textbf{Entity}                   & \textbf{Type} & \textbf{Explanation}                                                                                                                                \\ \hline
\textbf{Process} & \textbf{Target Process (TP)}                        &The normal running process in the system, which can be easily used by attackers to complete related attack operations. \\ \cline{2-3} 
                                  & \textbf{Malicious Process (MP)}                        & Attacker or malicious software, malicious file running process.                                                        \\ \hline
\textbf{File}    & \textbf{System Files (SF)}                        & Some important sensitive information files in the system.                                                                     \\ \cline{2-3} 
                                  & \textbf{Temporary high-value Files (TF)}                        & Temporary files created by attackers to save attack data or steal important information.                        \\ \cline{2-3} 
                                  & \textbf{Malicious Files (MF)}                        & Malicious code or script files deployed or written by attackers.                                                 \\ \hline
\textbf{Socket}                   & \textbf{Socket (SO)}                        & The programming interface used in network communication, such as url, domain name, ip address, etc.                                \\ \hline
\end{tabular}
}
\end{table*}

\begin{table}[h!t]
\setlength{\abovecaptionskip}{0pt}
\setlength{\belowcaptionskip}{-10pt}
\caption{The attribute of system events in \sn{}.}
\label{table:graph-relation}
\centering
\resizebox{\linewidth}{!}{%
\begin{tabular}{|c|c|}
\hline
\textbf{Subject and Object}   & \textbf{Relation Type}   \\ \hline
\textbf{ProcessToFile}    & read (RD), write (WR), execve (EX), chmod (CD), unlink (UK) \\ \hline
\textbf{ProcessToSocket}  & send to (ST), receive from (RF)                   \\ \hline
\textbf{ProcessToProcess} & fork (FR), inject (IJ), unlink (UK)       \\ \hline
\end{tabular}
}
\end{table}

\subsection{AsyncRAT \cite{AsyncRAT} Reports with Redundancy Filtered}\label{appendix:AsyncRAT}
The infections leverage process injection to evade detection by endpoint security software. The threat actor(s) behind these campaigns have been using 3LOSH to generate the obfuscated code responsible for the initial infection process. Infection processThe infection process begins with an ISO that contains a malicious VBScript that, when executed, initiates a multi-stage infection process. (VBS|ISO)\$Stage 1 Execution. The VBS contains junk data and uses string replacement to attempt to obfuscate the executed code. Once deobfuscated, the VBS execution is straightforward: It retrieves and executes the next stage from an attacker-controlled server. Stage 2 retrieval. As expected, the retrieved content is a PowerShell script passed to the Invoke-Expression (IEX) cmdlet and executed to continue the infection process. It is mainly responsible for creating a series of scripts that are executed and carry out various tasks needed for the malware to function. Across various samples analyzed, the directory locations and file names vary, but are functionally equivalent. First, the script checks for the existence of a directory at the following location:C:\textbackslash\textbackslash ProgramData\textbackslash\textbackslash Facebook\textbackslash\textbackslash System32\textbackslash\textbackslash Microsoft \textbackslash\textbackslash SystemData. If it doesn't already exist, the directory is created. The script then creates several additional scripts, writing content into each of them using a format similar to the following example. The following files are created in this manner: Office.bat, Office.vbs, Office.ps1, Microsofd.bat, Microsofd.vbs, Microsofd.ps1. All of these scripts are stored in the previously created directory. Finally, the Stage 2 PowerShell executes “Office.vbs” to begin the next step of the infection process. The first script, “Office.vbs” is executed by the Stage 2 PowerShell and invokes WScript to execute a batch file called “Office.bat” to continue the infection process. WScript batch file execution. This batch file, in turn, executes a PowerShell script called ‘Office.ps1’. PowerShell script execution. The next PowerShell script attempts to achieve persistence by creating a new Scheduled Task called “Office” that is executed immediately and then repeated every two minutes, as shown below. This scheduled task then executes “Microsofd.vbs” as part of the creation process. This next VBS initiates a short sleep before continuing. It is only responsible for executing “Microsofd.bat” to continue the infection process. This next batch file only contains a single line, which invokes PowerShell and executes ‘Microsofd.ps1’ PowerShell execution. This PowerShell script is the final script executed in this chain. This script contains two large GZIP blobs and another function responsible for decompressing them. This is accomplished by invoking aspnet\_compiler.exe, injecting the final payload, and executing it. The infection chain is more complicated, featuring the use of multiple script-based components (BAT, VBS, PS1) that facilitate the infection process. Additionally, while analyzing our original sample cluster and new samples created using the 3LOSH crypter, we identified several final payloads in both clusters that use the same infrastructure for post-compromise C2 communications. Matching RAT configs in both sample clusters. 

\begin{figure}[ht]
\centering
\includegraphics[width=\linewidth]{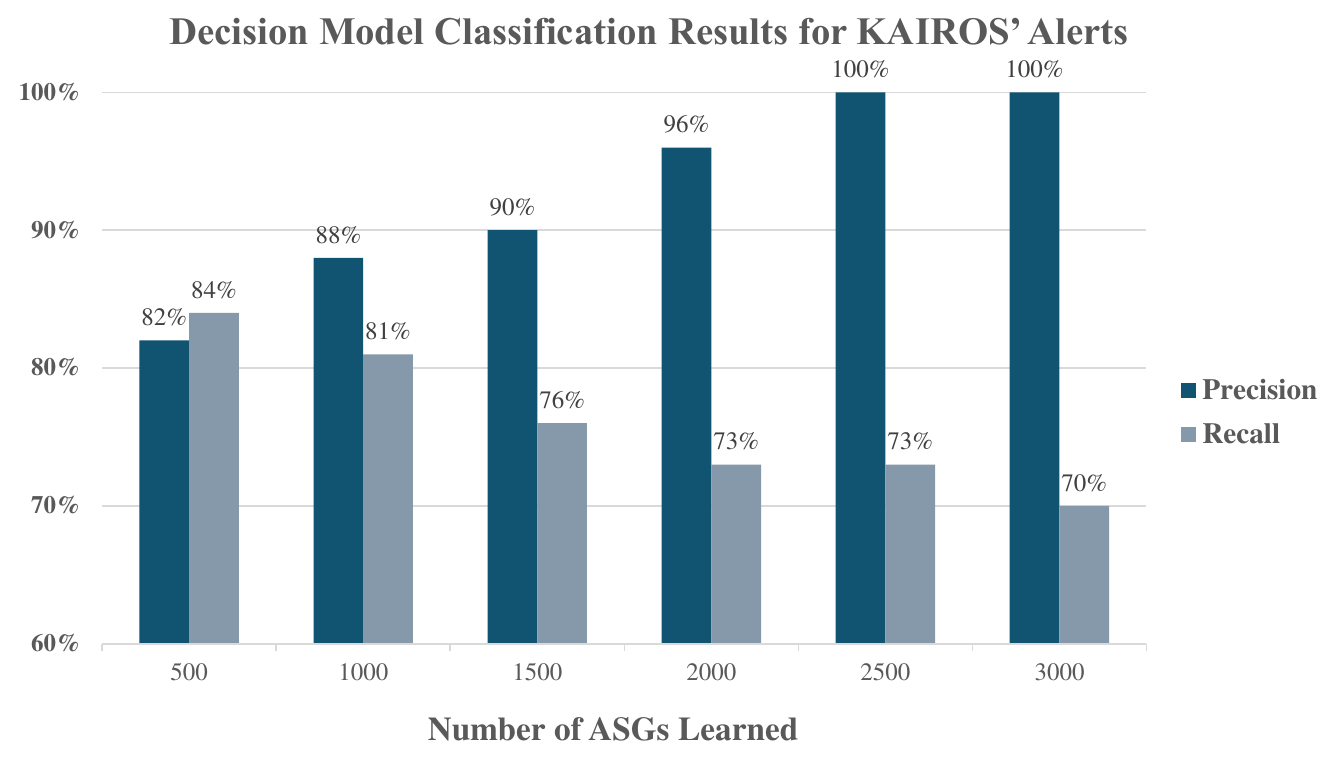}
\centering
\caption{Results of the decision (one-class) classification model’s determination of KARIOS’s 62 alarms (37 false positives/FPs and 25 true positives/TPs). Precision equals correctly identified FPs divided by model-identified FPs (in one-class classification model, TPs divided by sum of TP and FP), and recall equals correctly identified FPs divided by true FPs (37) in KAIROS (in one-class classification model, TPs divided by sum of TP and FN).}
\label{fig:eval5}
\vspace{-10pt}
\end{figure}

\subsection{The Usefulness of the ASGs Generated by \sn{} for Threat Detection}
\label{sec:eval_application}
To illustrate the effective application of \sn{} reconstruct ASGs in downstream tasks, we make decision boundaries for the anomaly detection model (take KARIOS \cite{cheng2023kairos} as an example) to help it remove false positives. We use phase-complete ASGs to train the decision model (one-class classification model), and the provenance graph reconstructed from the anomaly detection results is fed as an input to the decision model to eliminate false positives. Experimental results in Figure~\ref{fig:eval5} show that using the stage-complete ASGs training model decision model, we can reduce FPs in KARIOS by around 70\% (when the number of ASG reaches 2,500). We discuss this type of downstream tasks in the discussion section of this paper. In future work, we will conduct different research specifically on the downstream tasks of ASGs.
\par

\end{document}